\documentclass[final,1p,times]{elsarticle}
\usepackage{algorithmicx}
\usepackage{algorithm}
\usepackage{algpseudocode}
\usepackage{listings}  
\usepackage{graphicx}
\usepackage{subcaption}
\usepackage{pdflscape}

\usepackage{amssymb}
\usepackage{amsmath}
\usepackage{float}
\usepackage{amsthm}

\usepackage{xcolor}
\usepackage{pgfplots}
\pgfplotsset{width=10cm,compat=1.9}
\usepackage{ulem}
\usepackage{multirow}
\usepackage{tabu}

\biboptions{authoryear}

\journal{}

\begin{document}

\begin{frontmatter}

\title{A column-generation-based matheuristic for periodic train timetabling with integrated passenger routing}

\author[label1]{Bernardo Martin-Iradi\corref{cor1}}
\address[label1]{DTU Management, Technical University of Denmark, Akademivej Building 358, 2800 Kgs. Lyngby, Denmark}

\cortext[cor1]{Corresponding author}

\ead{bmair@dtu.dk}

\author[label1]{Stefan Ropke}
\ead{ropke@dtu.dk}


\begin{abstract}
{\footnotesize In this study, the periodic train timetabling problem is formulated using a time-space graph formulation.  
\footnotesize Three solution methods are proposed and compared
\footnotesize  where solutions are built by what we define as a \textit{dive-and-cut-and-price} procedure. An LP relaxed version of the problem with a subset of constraints is solved using column generation where each column corresponds to the train paths of a line. Violated constraints are added by separation and a heuristic process is applied to help to find integer solutions. The passenger travel time is computed based on a solution timetable 
and Benders' optimality cuts are generated
allowing the method to integrate the routing of the passengers.
\footnotesize We propose two large neighborhood search methods where the solution is iteratively destroyed and repaired into a new one and one random iterative method. 
\footnotesize The problem is tested on the morning rush hour period of the Regional and InterCity train network of Zealand, Denmark.
\footnotesize The solution approaches show robust performance in a variety of scenarios, being able to find good quality solutions in terms of travel time and path length relatively fast.
\footnotesize The inclusion of the proposed Benders' cuts provide stronger relaxations to the problem.
\footnotesize In addition, the graph formulation covers different real-life constraints and has the potential to easily be extended to accommodate more constraints.} 
\end{abstract}


\end{frontmatter}


\section{Introduction}
\label{sec1}

\begin{figure}[ht]
    \centering
    \includegraphics[width=0.85\textwidth]{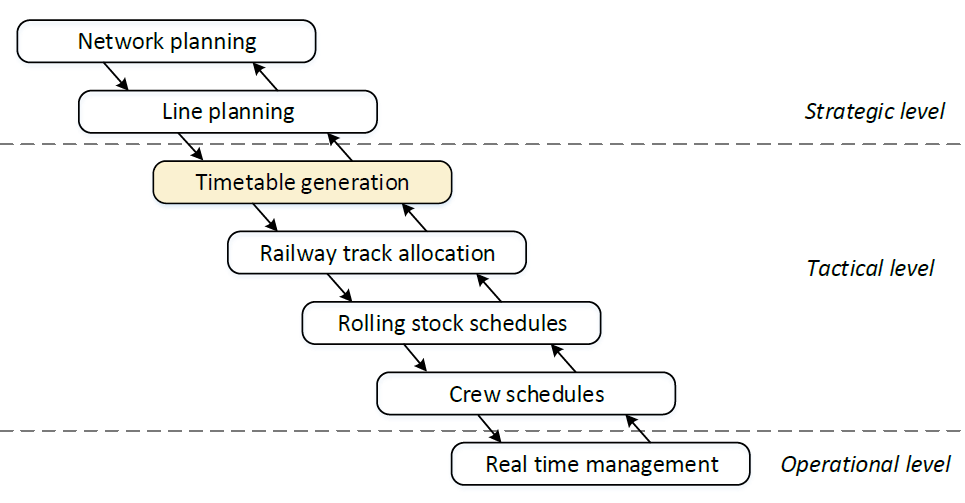}
    \caption{Railway planning process diagram adapted from \cite{lusby2011a}}
    \label{fig:RailwayPlanPhases}
\end{figure}

The planning process of railway companies is complex and is usually categorized into three main levels: \textit{strategic}, \textit{tactical} and \textit{operational}  \citep{Bussieck1997}. These levels form a hierarchical process used as a decision-making tool where each of the levels includes different problems whose solution is used as an input for the problems at the subsequent level as depicted in Figure \ref{fig:RailwayPlanPhases}. 

In this study, the focus is mainly on the generation of timetables which is at the tactical level of the planning process. 
For that, the network and lines running on it, decided at the strategical level, are assumed fixed.
The process of generating a timetable 
is formulated as the \textit{Train Timetabling Problem} (TTP) and its main goal is to determine the arrival and departure times at the stations for each of the train lines. 

The departure and arrival times are subjected to multiple track capacity constraints and specific requirements from the railway operating company. An obvious example of track capacity constraints is that two trains cannot be in the same track segment at the same time.
In order to avoid having two trains at the same track segment at the same time, a \textit{headway} is defined. The headway refers to the minimum time interval between two consecutive train movements and it is defined by the signaling system along the track. 
Likewise, a headway is defined for both departures and arrivals of consecutive trains along the same track segment. 
Moreover, 
a minimum dwell time is necessary to allow passengers to get on and off the train as well as changing drivers at specific stations. In the same way, minimum running times between two stations are limited by the train speed, acceleration, breaking capabilities and 
an additional buffer time 
also known as \textit{timetable margin}.

In general, the objectives are related to three main groups: customer satisfaction, robustness and cost-efficiency. These objectives may be conflicting in most cases. 
For instance, a timetable where all passengers have
direct connections to their destinations at a high frequency
would incur in an enormous operational cost for the train operating company (TOC).
Therefore, a compromise between conflicting objectives should be found.

\subsection{Focus of the paper}
In this study, we focus on the generation of timetables from the passengers' point of view while also analyzing the robustness of the solution. The model presented relies in two main assumptions: (1) the running times between stations are considered fixed and (2) the timetable should be symmetrical or close to symmetrical (we elaborate on this in Section \ref{sec:newGraph}). 
The main contributions of the paper are two-fold: We present (1) a new graph formulation that allows us to directly generate non-conflicting schedules for all the trains of a line and also to include additional operational constraints with minor adaptations and, 
(2) a Benders' decomposition formulation that enables the integration of passenger routing in the timetabling generation process.

\subsection{Paper structure}

Section \ref{sec:LR} lists several methods to solve the TTP through an extensive literature review. In Section \ref{sec:probForm} the model used and its characteristics are described. 
The solution methods used to solve the problem are described in Section \ref{Sec:SolM}, where each of the steps in the algorithms and how they interact together are carefully explained. 
Section \ref{sec:CaseStudy} introduces the case studied, summarizes the computational results obtained from different tests and conducts an analysis of them. 
The paper concludes in Section \ref{sec:conclusion} with a generic overview of the whole study and further study proposals.

\section{Literature review}\label{sec:LR}

The literature about train scheduling is extensive. The different publications apply a wide range of methods to different cases. Some of them consider just a corridor or a junction whereas others study a whole network. Moreover, the nature of the resulting timetable (i.e. periodic or non-periodic) also affects the algorithm proposed. Several extensive surveys have been published (see \cite{cordeau1998a}, \cite{CAPRARA2007129}, \cite{hansen2009railway}, \cite{lusby2011a}, \cite{cacchiani2012a} or \cite{harrod2012a}).

Most of the studies that model a network assuming the periodicity of the timetable (periodic timetable) are based on the Periodic Event Scheduling Problem (PESP) first introduced by \cite{Serafini:1989}.
\cite{odijk1996a} proposed a cutting plane algorithm to solve the PESP. Integer variables are used to ensure the travel intervals are respected and continuous variables to determine the arrival and departure times modulo the period.
Later, \cite{nachtigall1998periodic}, \cite{liebchen2002case} and \cite{Peeters03} studied the Cycle Periodicity Formulation (CPF) that leads to a significant speed up in the solution times compared to earlier models.
Given the effectiveness of the PESP, these models have been used to solve many network cases, whereas non-periodic approaches are used more often to model single-line corridors or congested networks where it may not be possible to schedule all trains in an efficient way.

\cite{szpigel1973a} presented one of the first Integer Linear Program (ILP) formulations for the non-periodic TTP. The formulation is regarded as a job-shop scheduling problem where jobs (trains) need to be assigned to machines (track segments). \cite{szpigel1973a} solved it using branch-and-bound applied to a Brazilian single-track line. \cite{jovanovic1991a} proposed a Mixed Integer Linear program (MILP) formulation where the arrival/departure times are defined with continuous variables and the order of trains with binary variables and tries to find a reliable timetable. 
\cite{Carey1995} proposed a mix of heuristic and branching procedure to solve a similar MILP as the one presented by \cite{jovanovic1991a} in a one-way corridor, and \cite{carey1994a} extended it to a two-way corridor showing that no additional constraints are needed.
In general, most of the models proposed for solving non-periodic timetables are used for scheduling multiple competing timetables from different operators.

Furthermore, \cite{brannlund1998a} introduced a pure ILP formulation where the time was discretized and therefore, the formulation could be represented as a graph where the nodes represent the arrival and departure time instants to each station. This new formulation is referred to as \textit{time-space graph} formulation but cannot be directly applied to large instances due to the large number of binary variables. As a result, further studying the LP relaxation of the model becomes more attractive and different methods have been developed based on it. The ILP formulation proposed by \cite{caprara2002a} defines a variable for each arc in the graph and it is solved using Lagrangian relaxation combined with sub-gradient optimization.  
\cite{Cacchiani2008} proposed a formulation where the variables refer to whole paths instead, and solved it applying column generation together with separation techniques. \cite{CACCHIANI2010Freight} extended the formulation presented by \cite{caprara2002a} to be applied in a network considering both passenger and freight trains and solved it using a similar procedure. \cite{MIN2011409} proposed a method for solving the train-conflict resolution problem with a column-generation based algorithm that takes advantage of the separability of the problem. Using a heuristic for the pricing problem (PP), the method is able to get near optimal conflict-free solutions in a few seconds. \cite{cacchiani2013a} applied dynamic programming to solve the clique constraints that arise in the graph formulations and developed an exact method whose performance is compared with various heuristics in \cite{CACCHIANI2010179}.
\cite{fischer2015a} formulates the TTP using a time-indexed graph and presents a method based on Lagrangian relaxation that improves the quality of the relaxation. \cite{fischer2017a} extends the approach to also allow overtaking possibilities.
\cite{zhou2017a} and \cite{zhang2019a} also take advantage of a graph formulation and effectively solve it using dual decomposition techniques. The methods are applied to the Beijing-Shanghai high speed corridor and show a better performance than the PESP model.

Last but not least, combining train timetabling and passenger routing has also been studied. \cite{kinder2008models} extended the PESP model to a time-space graph and implemented an iterative approach where the timetable is re-planned after doing passenger routing. \cite{gattermann2016integrating} present an integrated model that finds timetables and passenger routes in which passengers are distributed temporally using \textit{time-slices}. \cite{borndorfer2017passenger} also integrates timetabling and passenger routing in one model. The model tests and analyzes different passenger routing models on timetable optimization yielding significant improvements in travel time. 
\cite{farina2019phd} proposes a two-phase large neighborhood search heuristic for the combined train timetabling and passenger routing problem. The heuristic has similarities with the work presented in this paper, but employs different destroy and repair methods.
\cite{cacchiani2020} also implement a two-phase heuristic that aims at minimizing the passenger travel time. The method also accounts for the waiting time of the passengers at the stations and shows promising results in real-life instances.
Several studies also refer to the problem at hand as the \textit{demand-oriented train timetabling} problem.
\cite{li2017a} implements a mixed integer quadratic model for the dynamic version of the problem and shows that it can effectively reduce the total passenger travel time. \cite{zhou2019a} studies passengers' booking decisions instead of the classic queue principle and uses a two-level method which combines a bi-level programming model with a priority-based heuristic which also shows benefits in terms of travel time for passengers.

\subsection{Contribution and comparison to existing models}

The modeling approach used in this paper is based on the time-space graph proposed in \citet{caprara2002a}. As discussed in the literature review, this modeling approach has also been used in several later papers (e.g. \citet{Cacchiani2008} and \citet{CACCHIANI2010Freight}). In \citet{caprara2002a}, the integer programming model was solved using a Lagrangian relaxation heuristic. The Lagrangian subproblem solves a longest path problem through an acyclic network. In \citet{Cacchiani2008}, the problem was solved using column generation where the pricing problem also searches for longest paths through an acyclic network. We also solve the problem using column generation but use a pricing problem that can determine 1, 2 or 4 paths in one go. The pricing problem is solved as a standard shortest path problem (further details in Sections \ref{sec:probForm} and \ref{Sec:SolM}). This is possible due to tight frequency and symmetry constraints. There are several benefits of this approach: 1) The symmetry and frequency constraints are entirely handled in the pricing problem and fewer constraints are necessary in the master problem. 2) The LP relaxation produced by the master problem is potentially stronger compared to an approach that handles symmetry and frequency constraints in the master problem. 3) Fewer pricing problems must be solved. We believe that this is a major contribution of our paper. 

\citet{caprara2002a} already constructed a cyclic timetable. We use this as a basis to generate cyclic timetables with a one hour period, useful for modeling the passenger train timetabling problem that a train operator faces. Normally, this problem is solved using a PESP model and, to the best of our knowledge, it is the first time that the time-space graph approach is used for this application. 

The solution approach presented in this study constructs the
timetable while considering the routing of the passengers. 
A routing sub-problem is used to generate Benders' cuts that guide the model to optimize
the passenger travel time.
As the literature review shows, this is an emerging topic in passenger train timetabling and we believe that the paper at hand proposes a simple but useful approach for integrating the passenger routing with the train timetabling problem.

The method proposed in the paper at hand is based on work done in the master's thesis of Bernardo Martin-Iradi \citep{martin2018a}.

\section{Problem formulation} \label{sec:probForm}

The notation is based on the one from \cite{CACCHIANI2010Freight}. Let $S = \{1,...,s\}$ denote the set of stations in the network. 
The network can be represented as a mixed multi-graph $N = (S,E \cup A) $ where each vertex $i \in S$ represents a station in the network and each edge $e= (h,i)\in E$ represents a single-track segment between two stations with no intermediate stations in between that is used by trains traveling in both directions (i.e. from \textit{h} to \textit{i} and from \textit{i} to \textit{h}). Finally, each arc $a = (h,i) \in A$ represents a double-track segment between station \textit{i} and \textit{j} with no intermediate stations that can be used only by trains traveling in one direction (i.e. from \textit{h} to \textit{i}). The graph can contain multiple arc/edges connecting the same two stations. For instance, in the network here studied there are segments with four tracks between two same stations (two in each direction). Therefore, the adjacent stations in between can be connected with four arcs (two in each direction) in the multi-graph. 
For convenience, for each station $ i \in S$, let denote $\delta^+_{N}(i) \subseteq E \cup A$ the set of edges incident to $i$ and arcs leaving $i$, and $\delta^-_{N}(i) \subseteq E \cup A$ the set of edges incident to $i$ and arcs entering $i$. \\
Furthermore, for both mono and bi-directional tracks, minimum time intervals between departures/arrivals (i.e. headway) on the same track are required. Therefore, for each $e \in E \cup A$ and station $i$ of $e$, let denote:
 \begin{itemize}
     \item $d(i,e)$: minimum time interval between consecutive departures of trains traveling in the same direction from $i$ on the track segment $e$.
     \item $a(i,e)$: minimum time interval between consecutive arrivals of trains traveling in the same direction at $i$ on the track segment $e$.
 \end{itemize}
Moreover, in the case of single-tracks, additional time interval requirements need to be set for trains traveling in opposite directions. Therefore, for each edge $e \in E$ and station $i$ of $e$ where 
$i \in \hat{S}$ and $\hat{S}$ is the set of stations connected by single-track segments,
we denote:
\begin{itemize}
     \item $f(i,e)$: minimum time interval between an arrival at $i$ on $e$ and a departure from $i$ on $e$ of trains traveling in opposite directions.
     \item $g(i,e)$: minimum time interval between a departure from $i$ on $e$ and an arrival to $i$ on $e$ of trains traveling in opposite directions.
\end{itemize}
Furthermore, let $S^* \subseteq \hat{S}$ be the stations only connected by single-track segments. Therefore, for station $i \in S^*$ we define:
\begin{itemize}
     \item $h(i)$: minimum time interval between an arrival to $i$ and an arrival to $i$ of trains traveling in opposite directions.
\end{itemize}
In this case study, due to safety requirements, a minimum value of $d(i,e)$, $a(i,e)$ and $h(i)$ is defined,
whereas $f(i,e)=0$ and $g(i,e)$ is implicitly given by:
\begin{equation*}
     g(i,e) = \text{minimum travel time from } i \text{ to } h \text{ on } e + \text{minimum travel time from } h \text{ to } i \text{ on } e, 
 \end{equation*} 
where $h$ is the other endpoint of $e$.
The reason for $f(i,e) = 0$ is based on the rail infrastructure. At station $i$, each of the platforms has its own track and usually the length of the tracks until their merging point allows a train to depart on $e$ as soon as the other train has arrived from $e$. 

 \subsection{Lines and timetables notation}
\begin{figure}[ht]
 \centering
    \includegraphics[width=0.6\linewidth]{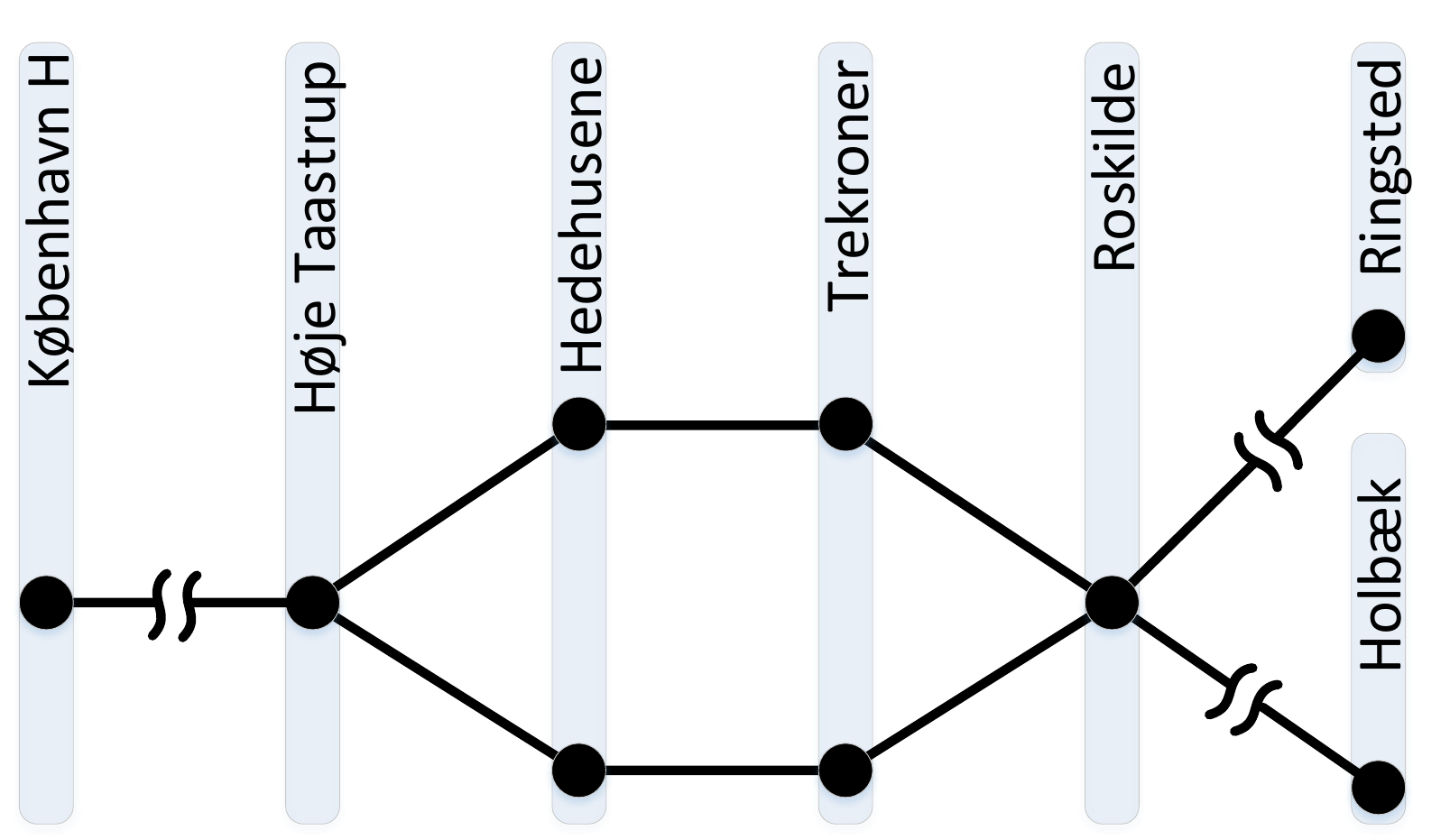}
    \caption{Illustration of the quadruple-track segment modelling in one direction.}
    \label{fig:quadTrackOneWay}
\end{figure}

 The different lines link two major stations with a number of intermediate stations in between. Let $L = \{1,...,l\}$ denote the number of operating lines in the network space and $D= \{1,2\}$ the direction of the line, $D=1$ for direction out of Copenhagen and $D=2$ for direction towards Copenhagen. Let $\Upsilon$ be the set of trains that cover the $L$ lines and $D$ directions. For each train $j  \in \Upsilon$ let denote $f_j$ the starting station and $e_j$ the ending station. Let $S^{j} : = \{ f_j,...,e_j\} \subseteq S$ be the ordered set of stations visited by train $j$ (stopping or not). Some segments between stations are formed by quadruple-track segments, meaning that each train can choose between two tracks to travel along that track segment. In this study, the quadruple-track segments connect various consecutive stations and it has been assumed that the train runs along the same track and cannot switch to the other track during the whole quadruple segment (see Figure \ref{fig:quadTrackOneWay}). Let $N^j = (S^j,A^j)$ be the auxiliary network for each train $j  \in \Upsilon$ where each arc in $A^j$ is either an arc in A or an edge in E with an orientation, corresponding to the unique travel direction of $j$ along the single-track. A timetable for each train is given by the departure time at $f_j$ and the arrival time at $e_j$, and the arrival and departure times for the intermediate stations 
 $S^j \backslash\{f_j,e_j\}$.
 Let $\phi_{j}(a)$ denote the running time along arc $a\in A^j$ of train $j \in \Upsilon$. Let $\omega^{min}_{j}(i)$ denote the minimum dwell time at station $i$ for train $j \in \Upsilon$ where $i \in S^{j} \setminus \{f_j,e_j\}$. In the same way, there is an upper bound in the dwell time (i.e. $\omega^{max}_{j}(i)$) in the form of an additional percentage of the minimum dwell time ($\omega^{max}_{j}(i) \propto \omega^{min}_{j}(i)$). Note that, for a line containing N stations, there are N-1 minimum running times and N-2 minimum dwell times defined in one direction. The mentioned parameters above are defined for each train meaning that the running and dwell time sets are defined independently for trains in different directions for the same line, as they may differ. Finally, the time horizon is defined as $T = \{1,...,t\}$ referring to a whole hour discretized into time instants of half a minute ($|T| = 120$ time instants) and each line has an associated running frequency $F^l$ indicating how many trains per hour cover each direction of that line. 
 
 \subsection{A graph representation} \label{sec:graphRep}
 \begin{figure}[ht]
     \centering
     \includegraphics[width=0.8\linewidth]{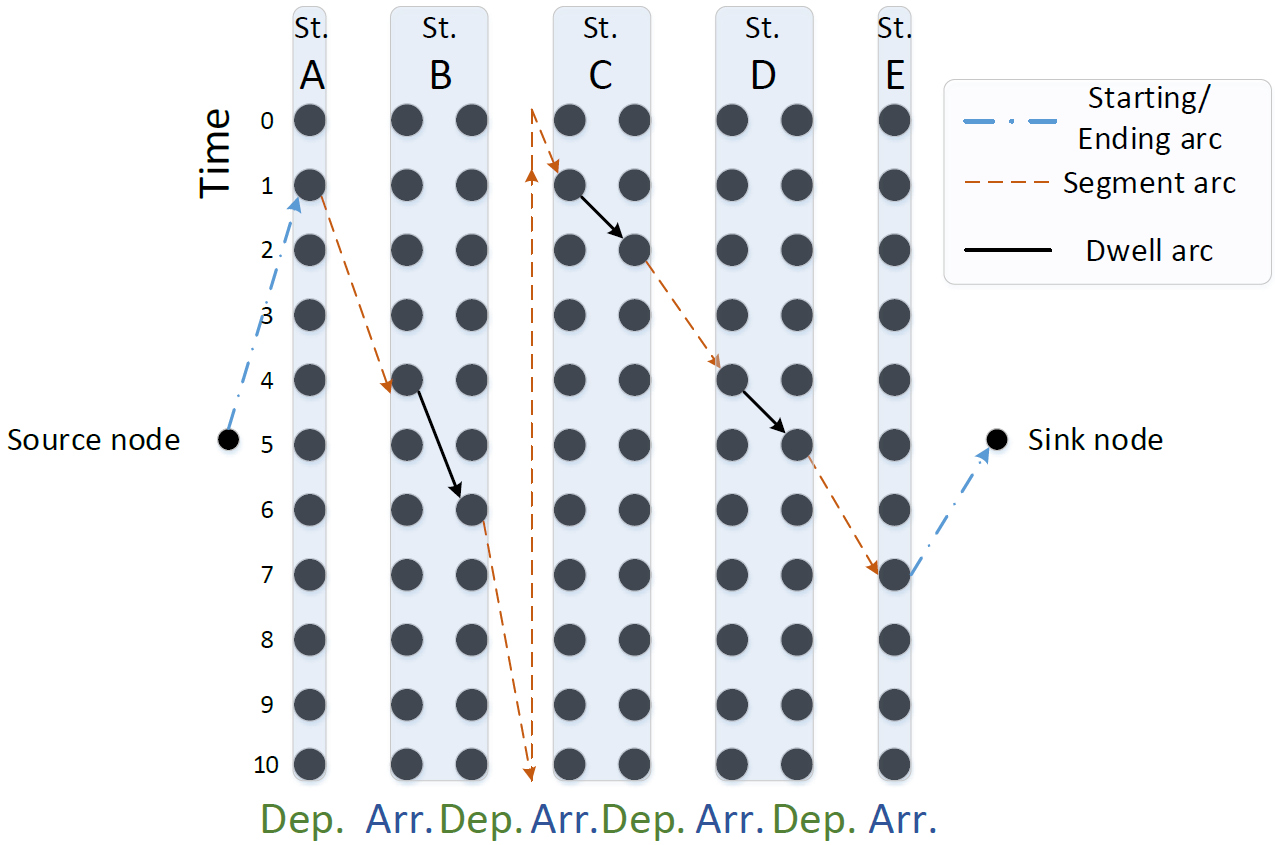}
     \caption{Graph representation of a train path with a time period of $|T|=10$. The nodes at $t=0$ correspond to a duplicate of the nodes of $t=|T|$.}
     \label{fig:graphPath}
 \end{figure}
 The problem can be defined using graphs to represent the possible timetables (from now on referred to as \textit{train paths}). 
 Let $G = (V,R)$ be a directed and acyclic space-time graph.
 A sub-graph $G^j = (V^j,R^j)$ can be defined for each train $j \in \Upsilon$ 
 (from now on referred to as \textit{Train graph}) in which the nodes represent the arrivals or departures at a station at a given time instant. Figure \ref{fig:graphPath} shows an example of a train path represented using a time-space graph.  
 
 The node set has the form 
 \begin{equation*}
     V^j = \{ \sigma^j, \tau^j \} \cup \bigcup_{a=\{h,i\} \in A^j} (U^a_i \cup W^a_h)
 \end{equation*}
 where $\sigma^j$ and $\tau^j$ are the \textit{artificial source node} and \textit{artificial sink node} respectively and the sets $W^a_{h}$ for $h\in S^{j} \setminus\{e_j\}$ and $U^a_{i}$ for $i \in S^{j} \setminus \{f_j\}$ represent the set of time instants where a train can depart from or arrive to station $h$ or $i$ on the track represented by arc $a\in A^j$ respectively (also called \textit{departure} and \textit{arrival} nodes). Let $u,w \in V^j$ be nodes of the node set and let $\theta(u)$ be the time instant associated with node $u$. Furthermore, let $\Delta(u,w) := \theta(w) - \theta(u)$ denote the time interval between nodes $u$ and $w$ if $\theta(w) \geq \theta(u)$ and $\Delta(u,w) := \theta(w) - \theta(u) + T$ otherwise. Due to the periodic nature of the time horizon $T$, it is said that node \textit{u precedes} or \textit{coincides} with node \textit{w} (i.e. $u \preceq w)$ if $\Delta(w,u) \geq \Delta(u,w)$ as it is assumed that all the travel times used in this study case are far from the time horizon of one hour. Table \ref{tab:DeltaTexample} illustrates the time interval calculation with one example.
 \begin{table}[ht]
\centering
\caption{Example of the time interval calculation between two nodes with a cycle time $|T|=60$}
\label{tab:DeltaTexample}
\begin{tabular}{cc|c}
\textbf{$\theta(u)$} & \textbf{$\theta(w)$} & \textbf{$\Delta(u,w)$} \\ \hline
10                & 15                & 5                   \\
15                & 10                & 55                 
\end{tabular}
\end{table}
 For convenience, for each station $ i \in S^j$, let denote $\delta^+_{N^j}(i) \subseteq A^j$ the set of edges incident to $i$ and arcs leaving $i$, and $\delta^-_{N^j}(i) \subseteq A^j$ the set of edges incident to $i$ and arcs entering $i$.
 The arc set $R^{j}$ for each graph can be defined by four main types of arcs.

     \textbf{Starting arc set:} These arcs connect the\textit{ artificial source node} with the set of nodes for the departure of the first station in the line.
    These arcs have a null cost (free arcs).
    
    \textbf{Segment arc set:} These arcs connect the nodes related to the departure time from one station to the nodes related to arrival time to the next station in the line.
     Furthermore, the arc needs to satisfy that $\Delta(w,u) = \phi_j(a) $ where $\phi_j(a)$ denote the travel time for arc $a\in A^j$. The cost of the arc corresponds to the \textit{travel time} between the departure and arrival instants in the respective sets.
     
     \textbf{Dwell arc set:} These arcs connect the nodes related to the arrival time to one station with the nodes related to departure time from the same station in the line.
     Furthermore, the arc needs to satisfy that $\Delta(u,w) \in [\omega^{min}_j(i),...,\omega^{max}_j(i)]$ for $i \in S^{j} \setminus \{f_j,e_j\}$. The cost of the arc corresponds to the \textit{dwell time} between the arrival and departure instants in the respective sets.
     
     \textbf{Ending arc set:} These arcs connect the set of nodes of the arrival to the last station in the line with the\textit{ artificial sink node}. 
    These arcs have a null cost (free arcs).

As a result, the timetable for train $j \in \Upsilon$ is defined by any path from the artificial source node $\sigma^j$ to the artificial sink node $\tau^j$.

\subsubsection{Main assumptions}
The final graph formulation presented in this study is based on the assumption that the travel time of each train along each track segment joining two stations is fixed. In other words, it is not possible to slow down the train along the track segment and, therefore, the departure time from one station uniquely determines the arrival time at the next station. Even if slowing down is something that has to be done at the operational level, this assumption is supported by the fact that, in practice, slowing down a train between two stations in most cases is equivalent to forcing the train to stop in an endpoint station of the track segment for a longer time and then to travel at the regular speed along the track. This statement is not true in general but it holds for realistic cases. In particular, experimental results performed by \cite{CAPRARA2006738} show that the solution values found by heuristic procedures are marginally affected by this additional constraint, whereas the corresponding running time per iteration is widely reduced, since the graph $G$ turns out to be much smaller (for each train, the number of segment arcs between two stations is equal to the number of departure nodes). Furthermore, the above assumption simplifies the mathematical representation of the problem, yielding simpler and stronger overtaking and crossing constraints (see sections \ref{Sec::OTcons} and \ref{Sec::Crosscons}).

Another characteristic of the model assumed is the need for a symmetric timetable. When the train services are identical in both running directions it is easier to plan the timetable since the train path in one direction uniquely defines the path of the train in the opposite direction. Therefore, symmetric timetables are easier to plan and are more attractive to passengers as 
same transfer times are provided between pairs of trains  in both directions
\citep{liebchen2006a}. Nevertheless, this type of timetable reduces the degrees of freedom in the planning process and it is more suitable when the passenger demands are similar in both directions.

As a result, these two main assumptions can lead to a new, more efficient, graph formulation. On one side, keeping the running times fixed reduces the number of nodes to half since the arrival of a train is directly defined by the previous departure. On the other side, assuming symmetric paths for each line requires just creating one train path for a line, as the remaining line train paths are automatically defined. 
 
 \subsection{Symmetric Line graph} \label{sec:newGraph}
 \begin{figure}[th]
    \centering
    \includegraphics[width=\linewidth]{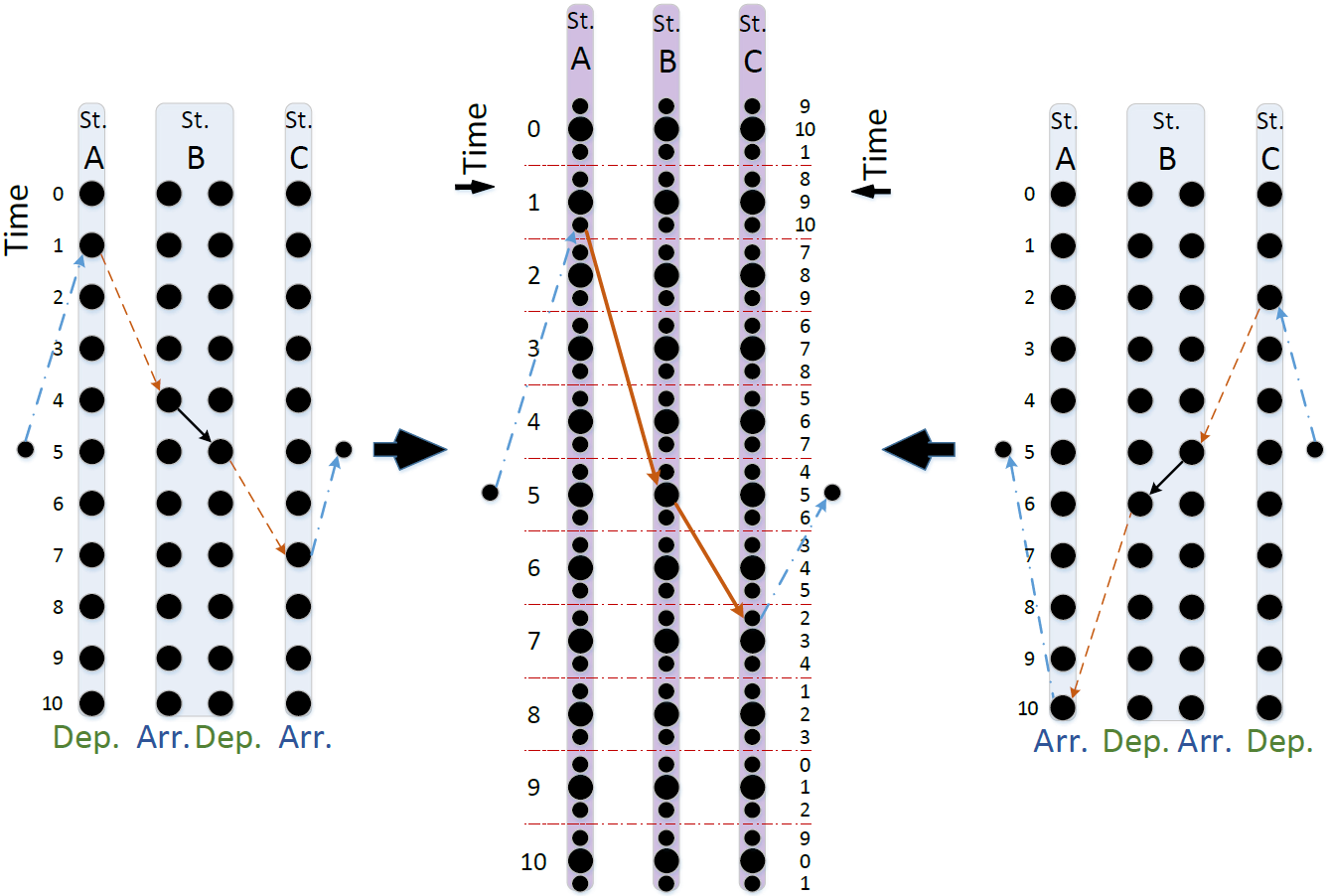}
    \caption{Representation of a path in the Symmetric Line graph as the combination of two paths in the respective Train graphs. In this example the symmetry gap is set to $\kappa = \pm 1$. 
    The time axis on the left for the Symmetric Line graph denotes the departure time instants for the left train and the time axis on the right denotes the arrival times of the right train. The nodes corresponding to $t=0$ are a duplicate of the nodes corresponding to $t=|T|$ and are added to help visualizing the symmetry of the paths in relation to the symmetry axis at $t=5$.
    }
    \label{fig:SymmExGraph}
\end{figure}

The Symmetric Line graph formulation is defined, as the name states, per each line instead of per train, meaning that fewer graphs are needed.
Ideally, each of the Symmetric Line graphs would include half of the nodes of one Train graph due to the fixed running times and symmetric paths. Nevertheless, in practice, 
due to the nature of the infrastructure, the running times in opposite directions for a given track segment are sometimes slightly different,
meaning that two exactly symmetrical paths cannot be achieved. Therefore, a \textit{maximum symmetry gap} $\kappa$ is considered.
A line is considered symmetrical, if, for each station, the departure time of the train in one direction and the arrival time of the train in the opposite direction sum to the period time $T$. The symmetry gap adds flexibility to this and allows to also consider the line to be symmetrical if the sums of departure and arrival times are within the interval (i.e. $|T| \pm \kappa$). Figure \ref{fig:SymmExGraph} shows an example of two trains of a line that are considered symmetrical and their corresponding path in the new proposed graph.
In this figure, the exactly symmetrical times at a station are depicted by larger nodes in the Symmetric Line graph and the symmetric instants that are within the gap considered ($\kappa$) are depicted with smaller nodes.
The primary time axis indicates the departures times of the left-to-right train and the secondary one indicates the arrival times of the right-to-left train.  
Starting with station A, the departure time of the left-to-right train is at time instant 1 and the arrival of the right-to-left train is at time instant 10. The sum of both times is 11 which is not equal to the planning horizon (10 in this case). Since the value is within the symmetry gap ($10 \pm 1$) it is symbolized with a small node.
For station B, the departure time of the left-to-right train is at time instant 5 and the arrival time of the right-to-left train is at time instant 5. The sum of both times is equal to 10 which is equal to the planning horizon meaning the departure and arrival of the trains are in perfect symmetry which is symbolized with the larger node.
Last, in station C, the arrival of the left-to-right train is at time 7 whereas the departure of the right-to-left train is at 2. The sum of both times sums to 9, which is not perfectly symmetrical but again lies within the symmetry gap ($10 \pm 1$) and therefore it is depicted as a small node.

\begin{figure}[ht]
    \centering
    \includegraphics[width=0.75\linewidth]{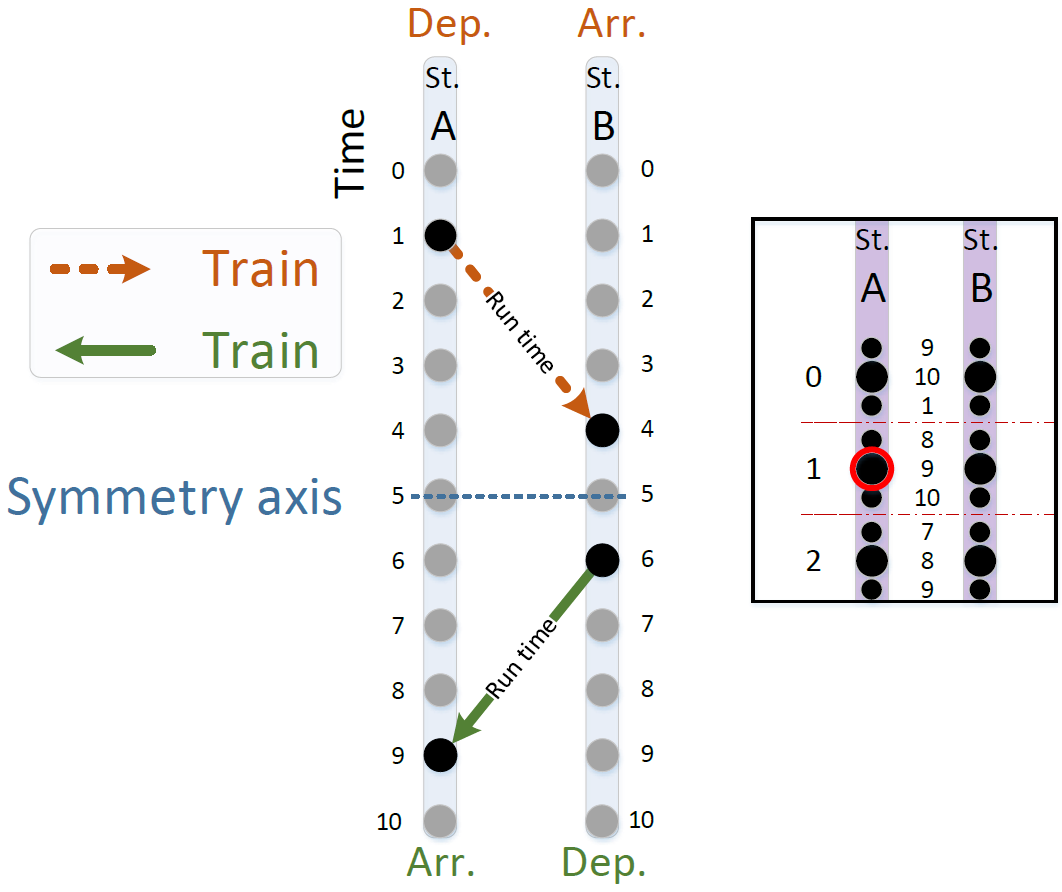}
    \caption[Representation of the Train graph nodes with one node (circled in red) of the Symmetric Line graph formulation]{Representation of the Train graph nodes associated with one node (circled in red) of the Symmetric Line graph formulation.
    Notice that by assuming fixed running times the departure time from A directly defines the arrival time at B and vice versa. This example also shows that the train paths are perfectly symmetrical with respect to the symmetry axis. The nodes corresponding to $t=0$ are a duplicate of the nodes corresponding to $t=|T|$ and are added to help visualizing the symmetry of the paths in relation to the symmetry axis at $t=5$.
    }
    \label{fig:oldNewnodes}
\end{figure}

Each node in the graph represents the departure and arrival times of two symmetrical train paths of the same line along a track segment. 
In other words, one node from the Symmetric Line graph notation is equivalent to four nodes of the Train graph notation (see Figure \ref{fig:oldNewnodes}).
As we increase the symmetry gap, the amount of symmetrical departure and arrival time combinations increases in accordance. Since each of those combinations can be seen as a node in the graph, the growth is translated in $(1+2|\kappa|)$ nodes per time instant and station.
 
 Regarding the arc set, the fact of assuming fixed running times allows us to merge the \textit{segment} and \textit{dwell} arc in a single \textit{segment+dwell} arc. The weight of these arcs is given by the sum of running and dwell time for both trains.
In order to avoid crossings or headway conflicts at a single-track segment, 
 all arcs that result in incompatible departures, arrivals or crossings are not included in the graph. This ensures that all paths in the new graph correspond to feasible and compatible train paths for the line.

Regarding lines using the quadruple track segments (see Figure 2), it is assumed that trains make the same choice of track in both directions.
 
The output of the Symmetric Line graph corresponds to a set of compatible train paths covering the line. Depending on the nature and frequency of the line, the amount may vary between one, two or four train paths, as explained below: 

If the line runs only during rush hour, trains only operate in one direction. This means that no symmetry is needed and a simple Train graph with fixed running times can be used. The output of it is just one train path, except if the frequency of the line is two trains per hour, then the output is two identical train paths exactly separated half an hour.

For regular lines, the output of the Symmetric Line Graph will be two symmetric train paths in opposite directions. If the frequency of the line is two trains per hour and direction, the output of the graph will correspond to two identical pairs of symmetric train paths separated by half an hour.

 \subsection{ILP formulation}\label{sec:ilp}
 In this section, the model is formulated as an ILP. In order to illustrate the different parts of the formulation, the notation of the Train graph is used. As it is explained in Section \ref{sec:newGraph}, the set of nodes of the Symmetric Line graph are formed by combinations of node sets from the Train graph formulation. 
 
 \subsubsection{Formulation without track capacity constraints}
 The problem can be formulated as a version of the Set Packing Problem (SPP) that aims to minimize the sum of total path lengths. The binary variable $\lambda_q \in \{0,1\}$, $q \in Q$ defines if the group of line paths $q$ is included in the optimal solution where $Q$ is the set of possible line group paths. The parameter $c_q$ denotes the cost of choosing the group of line paths $q \in Q$ that is the sum of path lengths.
 The formulation without the track capacity constraints is stated as follows:
 \begin{equation} \label{eq::objfSP}
min \ \ \ \sum_{q \in Q} c_q \cdot \lambda_q
\end{equation}
\hspace{3cm}\textbf{s.t.}
\begin{equation} \label{cons::SPcL}
\sum_{q \in Q^l} \lambda_{q}=1  \ \ \ \ \forall l \in L
\end{equation}
\begin{equation} \label{cons::SPbin}
    \lambda_q \in \{0,1\} \ \ \ \ \forall q \in Q
\end{equation}
The objective function minimizes the cost (path lengths) of the solution train paths. Constraints (\ref{cons::SPcL}) ensure that train paths are chosen to cover each line where $Q^l$ is the set of possible line group paths for line $l \in L$ and constraints (\ref{cons::SPbin}) state the binary property of the decision variable.

\subsubsection{Headway constraints}
Headway constraints are one of the track capacity constraints and ensure the minimum headway times between consecutive arrivals and departures at stations in the network.
\begin{equation} \label{cons::SPcA}
\sum_{\begin{array}{c}
v\in U_{i}^{a}:v\preceq u\\
\Delta(v,u)<a(i,a)\\
\\
\end{array}}\sum_{\begin{array}{c}
q\in Q_{v}\\
\\
\end{array}}\lambda_{q}\leq1,i\in S,a\in\delta_{N}^{-}(i),u\in U_{i}^{a},
\end{equation}
\begin{equation} \label{cons::SPcD}
\sum_{\begin{array}{c}
v\in W_{i}^{a}:v\preceq w\\
\Delta(v,w)<d(i,a)\\
\\
\end{array}}\sum_{\begin{array}{c}
q\in Q_{v}\\
\\
\end{array}}\lambda_{q}\leq1,i\in S,a\in\delta_{N}^{+}(i),w\in W_{i}^{a},
\end{equation}
\begin{equation} \label{cons::SPcAS}
\sum_{\begin{array}{c}
e\in\delta_{N}^{-}(i)\cap E\\
\\
\end{array}}\sum_{\begin{array}{c}
v,u\in U_{i}^{e}:v\preceq u\\
\Delta(v,u)<h(i,e)\\
\theta(u)=t
\end{array}}\sum_{\begin{array}{c}
q\in Q_{v}\\
\\
\end{array}}\lambda_{q}\leq1,i\in \hat{S}^*,t\in T,
\end{equation}
Let $Q_v$ be the set of line group paths that use node $v$. Constraints (\ref{cons::SPcA}) and (\ref{cons::SPcD}) enforce that the minimum headway distance between consecutive arrivals and departures at each station respectively, of trains in the same direction, is respected. Moreover, constraints (\ref{cons::SPcAS}) ensure that in the single-track segments the minimum headway between trains arriving in opposite directions is respected. 

\subsubsection{Overtaking constraints} \label{Sec::OTcons}

\begin{figure}[ht]
    \centering
    \includegraphics[width=0.8\linewidth]{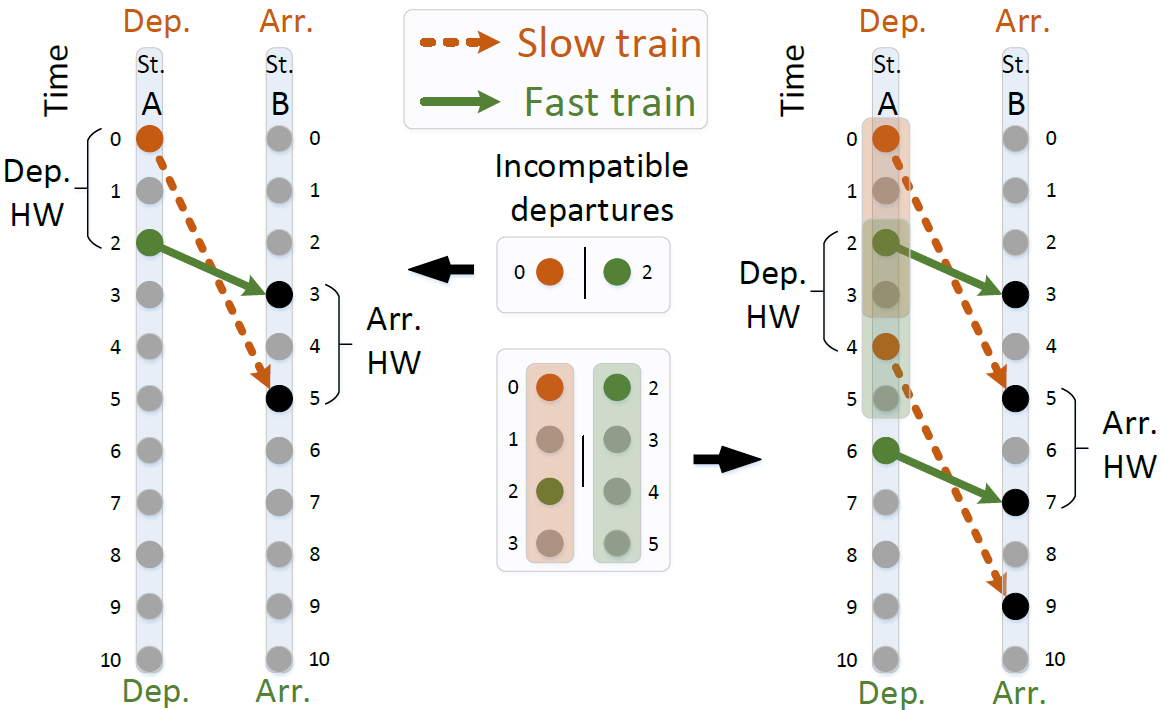}
    \caption[Illustration of an overtaking constraint]{Illustration of an overtaking where $a(h,a) = 2$ and $d(i,a) = 2$. The left one is the simple version of the constraint while the right one is the stronger version implemented in this study.}
    \label{fig:SimpleOTCons}
\end{figure}

It is not allowed that two trains traveling in the same direction on the same track overtake each other.

A basic example of an overtaking is shown on the left side of Figure \ref{fig:SimpleOTCons} where both train departures are incompatible. The basic overtaking constraint would enforce that, at most, one slow train will depart from $t=0$ or one fast train will depart from $t=2$. In this study, a stronger version of this basic constraint is formulated based on the ones from \cite{CACCHIANI2010Freight}.

The following constraints (\ref{cons::SPcOT}) are defined for every pair of trains $j,k$ along an edge/arc $a=(i,h)$ that is an arc in both auxiliary networks $N^{j}$ and $N^{k}$. Moreover, $j$ is considered the "slow" train and $k$ is the train that can actually overtake it. Therefore, the travel time of train $j$ should be greater than the one from train $k$ (i.e. $\phi^j(a) > \phi^k(a)$). For a constraint, we define an earliest possible departure from $i$ for trains $j$ and $k$. These departure nodes are denoted $v_1$ and $v_2$ respectively. Node $v_1 \in W^a_i\cap V^{j}$ and node $v_2 \in W^a_i\cap V^{k}$ correspond to departure nodes that are incompatible with each other (i.e. if train $j$ departs at $\theta(v_1)$, then train $k$ cannot depart at $\theta(v_2)$ and vice versa). The two trains $j,k$ are considered incompatible when either min$\{ \Delta(v_1,v_2),\Delta(v_2,v_1)\} < d(i,a)$, meaning that their departures are too close in time or min$\{ \Delta(u_1,u_2),\Delta(u_2,u_1)\} < a(i,a)$ where $u_1,u_2$ are the respective arrival nodes for $j,k$ corresponding to $v_1,v_2$, meaning that their arrivals to the next station are too close in time or $v_1 \prec v_2 \prec u_2 \prec u_1$ meaning that train $k$ overtakes train $j$ along the track.\\
Then, 
$v_3 \in W^a_i\cap V^{j}$ 
can be defined as the earliest possible departure of train $j$ that is compatible with $\theta(v_2)$ such that $v_1 \prec v_3$. Analogously, 
$v_4 \in W^a_i\cap V^{k}$
can be defined as the earliest possible departure of train $k$ that is compatible with $\theta(v_1)$ such that $v_2 \prec v_4$. It can be seen that any departure of train $j$ from $[v_1,v_3)$ is incompatible with any departure of train $k$ from $[v_2,v_4)$. 

This stronger version of the constraint is illustrated in the right side of Figure \ref{fig:SimpleOTCons}. Let $Q_{w}^{l^{j}}$ be the set of line group paths that use node $w$ and belong to train $j$ of line $l$. Nodes $v_1$ and $v_3$ are depicted as the first and second slow train nodes in time respectively and nodes $v_2$ and $v_4$ are depicted as the first and second fast train nodes in time respectively. Note that in the illustration the minimum departure and arrival headway ($a(i,e)$ and $d(i,e)$) are respected for the trains but they overtake each other along the track.

\begin{multline} \label{cons::SPcOT}
\sum_{\begin{array}{c}
w\in W_{i}^{a}\cap V^{j}:\\
v_{1}\preceq w\preceq v_{3}\\
\\
\end{array}}\sum_{\begin{array}{c}
q\in Q_{w}^{l^{j}}\\
\\
\end{array}}\lambda_{q}+\sum_{\begin{array}{c}
w\in W_{i}^{a}\cap V^{k}:\\
v_{2}\preceq w\preceq v_{4}\\
\\
\end{array}}\sum_{\begin{array}{c}
q\in Q_{w}^{l^{k}}\\
\\
\end{array}}\lambda_{q}\leq1,\forall j,k\in\Upsilon,v_{1},v_{2}\in W_{i}^{a}, \\
    (\text{where } l^j \neq l^k, d^j=d^k,  \ i,h \in S^{j} \cap S^{k}, a=(i,h)\in(A^{j}\cap A^{k}) ) 
\end{multline}

\subsubsection{Crossing constraints} \label{Sec::Crosscons}

\begin{figure}[ht]
    \centering
    \includegraphics[width=0.8\linewidth]{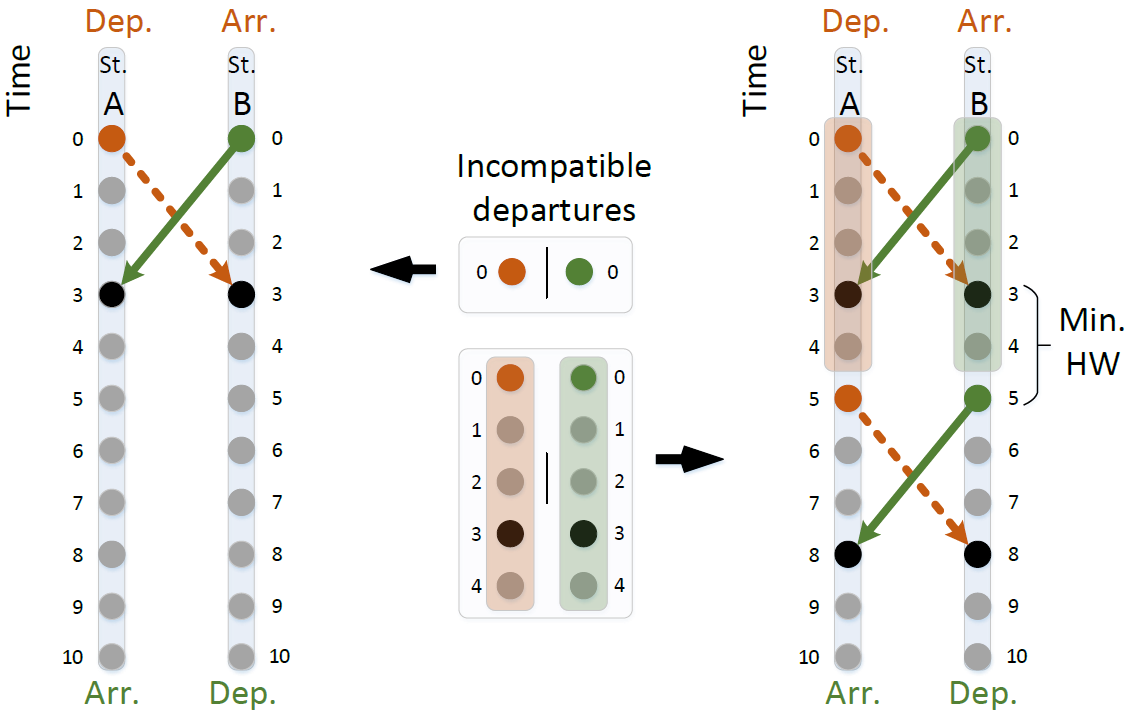}
    \caption[Illustration of a Crossing constraint]{Illustration of a crossing where $f(h,e) = 2$. The left one is the simple version of the constraint while the right one is the stronger version implemented in this study.}
    \label{fig:SimpleCrossCons}
\end{figure}

It is not allowed that two trains traveling in opposite directions are on the same single-track segment at the same time.

A basic example of a crossing is shown on the left side of Figure \ref{fig:SimpleCrossCons} where both departures are incompatible. The basic constraint corresponding to this crossing would enforce that, at most, one slow or fast train will depart from $t=0$. In this study, a stronger version of this basic constraint is formulated based on the ones from \cite{CACCHIANI2010Freight}.

The following constraints (\ref{cons::SPcross}) are defined in a similar way to constraints (\ref{cons::SPcOT}). They are defined for every pair of trains $j,k$ traveling in opposite directions such that $e=(i,h)$ and $(h,i)$ are arcs in the auxiliary networks $N^{j}$ and $N^{k}$ respectively and correspond to the set of edges $E$ in the network. For a constraint, we define an earliest possible departure from $i$ and $h$ for trains $j$ and $k$ respectively. These departure nodes are denoted $v_1$ and $v_2$ respectively. Node $v_1 \in W^e_i\cap V^{j}$ and node $v_2 \in W^e_h\cap V^{k}$ correspond to departure nodes that are incompatible with each other (e.g. if train $j$ departs at $\theta(v_1)$, then train $k$ cannot depart at $\theta(v_2)$ and vice versa). The two trains $j,k$ are considered incompatible when either $ u_2 \preceq v_1$ and $ \Delta(u_2,v_1) < f(i,e)$ or $ u_1 \preceq v_2$ and $ \Delta(u_1,v_2) < f(i,e)$, meaning that arrival to and departure from the same station are too close in time or $v_1 \prec u_2$ and $ \prec v_2 \prec u_1$ meaning that train $j$ and train $k$ cross each other along the track.\\
Then, $v_3 \in W^e_i\cap V^{j}$ can be defined as the earliest possible departure of train $j$ that is compatible with $\theta(v_2)$ such that $v_1 \prec v_3$. Analogously, $v_4 \in W^e_h\cap V^{k}$ can be defined as the earliest possible departure of train $k$ that is compatible with $\theta(v_1)$ such that $v_2 \prec v_4$. It can be seen that any departure of train $j$ from $[v_1,v_3)$ is incompatible with any departure of train $k$ from $[v_2,v_4)$. 

This stronger version of the constraint is illustrated in the right side of Figure \ref{fig:SimpleCrossCons}. Nodes $v_1$ and $v_3$ are depicted as the first and second slow train nodes in time respectively and nodes $v_2$ and $v_4$ are depicted as the first and second fast train nodes in time respectively.
Note that even if the minimum arrival headway ($f(h,e)$) is respected by the trains departing, they cross each other along the track.

\begin{multline} \label{cons::SPcross}
\sum_{\begin{array}{c}
w\in W_{i}^{e}\cap V^{j}:\\
v_{1}\preceq w\preceq v_{3}\\
\\
\end{array}}\sum_{\begin{array}{c}
q\in Q_{w}^{l^{j}}\\
\\
\end{array}}\lambda_{q}+\sum_{\begin{array}{c}
w\in W_{i}^{e}\cap V^{k}:\\
v_{2}\preceq w\preceq v_{4}\\
\\
\end{array}}\sum_{\begin{array}{c}
q\in Q_{w}^{l^{k}}\\
\\
\end{array}}\lambda_{q}\leq1,\forall j,k\in\Upsilon,v_{1}\in W_{i}^{e},v_{2}\in W_{h}^{e} \\
    (\text{where } l^j \neq l^k,d^j\neq d^k, \ i,h \in S^{j} \cap S^{k}, e=(i,h)\in E^{j}, (h,i) \in E^{k}) 
\end{multline}

\subsubsection{
Sibling
constraints} \label{sec:freqCons}
There are specific pairs of lines that share identical or similar first and last stations but have slightly different stopping patterns.
These pairs of lines (from now on referred to as \textit{
sibling 
lines}) should be spread along the cycle time as much as possible. In order to do so, the 
sibling 
constraints behave in the same way as the departure headway constraints (\ref{cons::SPcD}). Let $T_s$ denote the minimum time interval between consecutive departures of sibling lines in one direction at each station. Finally let $\Xi := \{(m_1,n_1),...,(m_k,m_k)\}$ denote the set of 
sibling 
line pairs along the network where $m_k,n_k \in L$.

\begin{multline}\label{cons::SPSib}
    \sum_{\begin{array}{c}
v\in W_{i}^{a}:v\preceq w\\
\Delta(v,w)<T_{s}\\
\\
\end{array}}\sum_{\begin{array}{c}
q\in\{Q_{v}^{l^{j}},Q_{v}^{l^{k}}\}\\
\\
\end{array}}\lambda_{q}\leq1,\forall(l^{j},l^{k})\in\Xi,d\in D,w\in W_{i}^{a}, \\
    (\text{where } j,k \in \Upsilon, i \in S^{j}\cap S^{k}, a \in \delta^+_N(i) \cap (A^{j}\cap A^{k})
\end{multline}
Constraints (\ref{cons::SPSib}) ensure that all the departures of
sibling 
lines from any common station are spread at least a time interval of $T_s$ in each direction.

\subsection{Passenger routing model formulation}\label{sec:passForm}
In order to route the passengers between stations, we introduce a multi-commodity flow problem (MCFP) formulation which is integrated with the ILP formulation by using a timetable solution as input information. 
Let $\bar{G} = (\bar{V},\bar{A}) $ be a graph formed by the set of nodes $\bar{V}$ and set of arcs $\bar{A}$. 
There is a node for each line $l\in L$, station $s\in S$ and time $t \in T$.
Let $K$ be the set of commodities. We define each pair of \textit{origin-destination} stations as a commodity $k \in K$ and the demand travelling between the corresponding origin and destination stations is given by an origin-destination ($OD$) matrix. Additionally, there is an artificial source and sink node $o_k, d_k$ per commodity $k \in K$.
The set $\bar{A}$ of passenger flow arcs is formed by different subsets:
\begin{itemize}
    \item $\bar{A}_r \subseteq \bar{A}$: \textit{Timetabling} arcs. Set of arcs corresponding to riding a timetabled train between consecutive stations. Due to the fixed running times, there are $T$ arcs between consecutive stations per line and direction.
    \item $\bar{A}_d \subseteq \bar{A}$: \textit{Dwell} arcs. Set of arcs corresponding to waiting time at a station, either dwelling on the train or waiting for the train to transfer to. There is one arc connecting two consecutive time instants in each station.
    \item $\bar{A}_s \subseteq \bar{A}$: \textit{Source} and \textit{sink} arcs. Set of arcs leaving one of the artificial source nodes $o_k$ or entering one of the artificial sink nodes $d_k$. For any origin station $i$, the artificial source nodes of commodities having station $i$ as \textit{origin}, are connected with the departures of trains stopping at station $i$. Likewise, all the arrivals of trains stopping at station $i$ are connected with the sink node of commodities that have station $i$ as \textit{destination}. 
    \item $\bar{A}_t \subseteq \bar{A}$: \textit{Transfer} arcs. Set of arcs to transfer between pairs of lines at a common station. 
    For each node at a station with transfer options, there is one transfer arc to each train belonging to different lines, that also visit the station.
\end{itemize}

\begin{figure}[ht]
    \centering
    \includegraphics[width=0.9\textwidth]{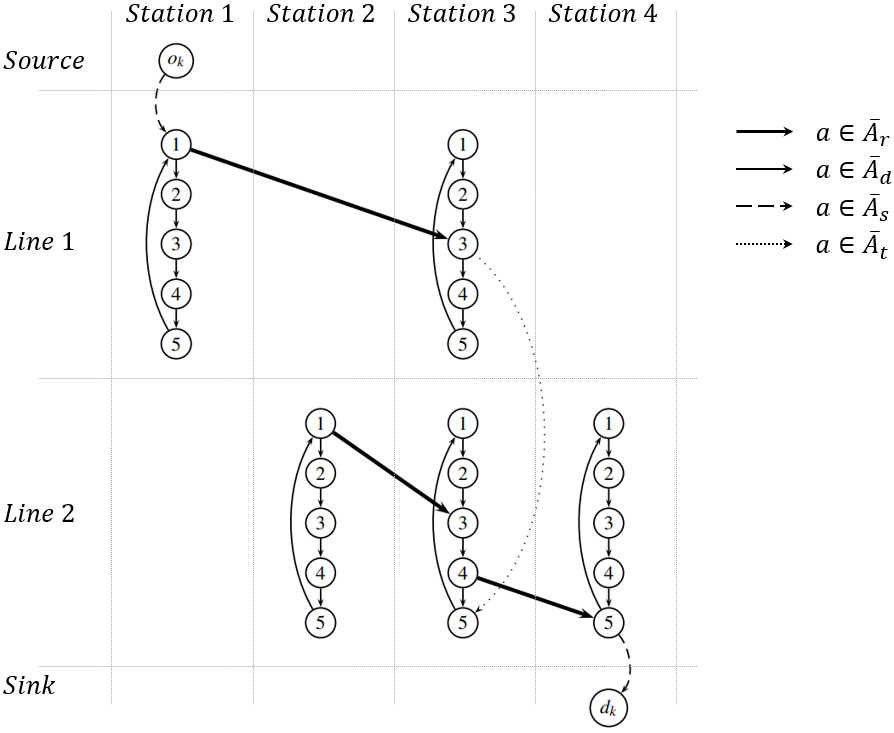}
    \caption[Passenger routing graph example]{Passenger routing graph example with a cycle period of $|T|=5$. In this case a subset of nodes and arcs is represented for trains of two different lines covering a subset of stations. The number in the node indicates a time instant. Only the artifical source and sink nodes for one commodity $k \in K$ are shown, which in this case, is the pair of stations $(1,4)$. }
    \label{fig:passGraphEx}
\end{figure}
A small example of the different elements in the graph are shown in Figure \ref{fig:passGraphEx}. 
We identify a possible routing path for passengers travelling from station 1 and 4 (i.e. from $o_k$ to $s_k$). This path consists on (1) boarding a train from line 1 that departs at time 1 from station 1, (2) getting off at station 3 and transferring to a train of line 2 that departs at time 4 and, (3) getting off at station 4 at time 5. Notice that the, in this example, we consider a minimum transfer time of 2, meaning that the transfer arc will allow us to board a train at time 5 at earliest.

Let $f_a^k$ be a variable that states if arc $a\in \bar{A}$ is used for commodity $k \in K$ and let $t_a$ be the time to traverse arc $a\in \bar{A}$. Finally, let $x_a$ be a variable that defines if timetabling arc $a \in \bar{A}_r$ can be used. These variables refer to the timetable solution given and are kept constant in this problem.
Notice that the formulation could be easily integrated with (\ref{eq::objfSP})-(\ref{cons::SPSib}) by using a constraint that maps the $\lambda_q$ variables to $x_a$. In order to avoid additional mathematical notation, this step has not been included. 

The problem is formulated as follows:

\begin{align}
    \min &  \sum_{k \in K} OD_k \sum_{a \in \bar{A}} t_a f^k_a \label{mcfp:obj} \\
    \sum_{a \in \delta^+(o_k)} f^k_a = 1 \quad &  \forall k \in K \label{mcfp:orig} \\
    \sum_{a \in \delta^-(d_k)} -f^k_a = -1 \quad & \forall k \in K \label{mcfp:dest} \\
    \sum_{a \in \delta^+(v)} f^k_a - \sum_{a \in \delta^-(v)} f^k_a = 0 \quad & \forall k \in K, v \in \bar{V}\backslash\{o_k,d_k\} \label{mcfp:flow} \\
    f_a^{k}- x_{a} \leq 0 \quad & \forall k \in K, a\in \bar{A}_r \label{mcfp:allow} \\
    f_a^k \geq 0 \quad & \forall k\in K, a \in \bar{A} \label{mcfp:pos}
\end{align}

The objective function (\ref{mcfp:obj}) minimizes the travel time of all passengers. Constraints (\ref{mcfp:orig}) and (\ref{mcfp:dest}) ensure that one arc is leaving from the source node and arriving to the sink node respectively for each commodity. 
Constraints (\ref{mcfp:flow}) ensure the flow conservation and constraints (\ref{mcfp:allow}) only allow to use arcs enabled by a timetable solution. Finally, constraints (\ref{mcfp:pos}) define the variables as linear positive. 
Notice that the capacity of the arcs is not limited, meaning that all passengers can board the same train. According to DSB (the train operator of the network studied), this is a fair assumption for this case.
This method is based on studies such as the ones proposed by \cite{schbel_et_al:OASIcs:2006:660} and \cite{rezanova2015line}.
The problem can be decomposed into $K$ different sub-problems, one per commodity and the totally unimodular structure of the problem formulation allows us to obtain an integer optimal solution by solving the LP model.

\section{Solution method} \label{Sec:SolM}
Three solution methods are presented are based on 
what we call, a \textit{dive-and-cut-and-price} procedure that heuristically solves the ILP formulation presented in Section \ref{sec:ilp}.
A Restricted Master Problem (RMP) is initialized with a subset of rows. Promising columns and violated cuts are added to it by column generation and separation procedure respectively in order to find an optimal LP solution. Then, branching is enforced through a dive heuristic in order to achieve integrality. Finally, the passengers are routed using the solution timetable and the travel time computed by solving
(\ref{mcfp:obj})-(\ref{mcfp:pos}).

Two of the methods are based on a large neighborhood search that iteratively transforms the solution by partially destroying and re-building it again. One of them uses the MCFP as a sub-problem to generate Benders' cuts for the RMP, helping to further integrate the passenger routing.
The third method is a simple iterative process where a solution is fully constructed at every iteration.

Each of the steps in the methods is explained in detail in the following sections.

\subsection{Column generation procedure} \label{sec:CG}
Taking into account the cycle time, the size of the network and the symmetry gap allowed, the number of possible line train paths to be considered is extremely large. In order to handle that amount of variables efficiently, column generation techniques are necessary.

A reduced version of the Master Problem (MP) is initially considered known as the Restricted Master Problem (RMP) that includes only a subset of the variables. These initial variables can just be a set of "dummy" artificial variables that satisfy the constraints of the RMP. 
For each line $l \in L$ a pricing problem is created (i.e. $PP^l$) that is in charge of providing line paths objects ($q\in Q^l$) that can potentially improve the current solution.

The formulation of the RMP is identical to the one of the original problem (see constraints (\ref{eq::objfSP})-(\ref{cons::SPSib})) except for the relaxed version of the decision variable (constraint (\ref{cons::RMPlinear})). 
\begin{equation} \label{cons::RMPlinear}
    \lambda_q \geq 0 \ \ \ \ \forall q \in Q
\end{equation}

\subsubsection{Pricing Problem}
The goal of the PP is to find new promising train paths for the RMP. There is one PP per line and their function is to create a group of \textit{line train paths} (referred to as a \textit{column}) with the potential to improve the objective function. 
Here is where the Symmetric Line graph formulation described in section \ref{sec:newGraph} becomes relevant.
The use of a single graph for all the train paths of a line reduces the PP to a single \textit{shortest path problem}.
It can be noticed that the dual value of constraints (i.e. (\ref{cons::SPcA}) - (\ref{cons::SPSib})) can be subtracted on the edge weights. Since, they are non-positive, we guarantee that the graph has always non-negative edge weights.
Therefore, and knowing that the graph is directed acyclic (see Section \ref{sec:newGraph}), this problem can be solved using
a dynamic programming algorithm.

Finally, to compute the reduced cost of a given path we need to subtract the dual value of constraint (\ref{cons::SPcL}) for the given line, which is a real number and can lead to a final negative reduced cost.
Every time the PP finds a column $q \in Q^l$ with a negative reduced cost, it is added as a new variable to the RMP and it is included in all the constraints where it has a non-zero coefficient.

\subsection{Separation procedure} \label{sec:separation}
It is decided to add Constraints (\ref{cons::SPcOT})-(\ref{cons::SPSib}) by separation as the total amount is too large and only a reduced amount of them may be binding. The headway constraints are considered from the beginning in order to provide guidance to the column generation process.

Once the column generation procedure stops providing columns with negative reduced cost the separation procedure is applied. The separation of constraints (\ref{cons::SPcOT})-(\ref{cons::SPSib}) is done by enumeration and are checked in the same iteration. Every constraint that is violated by the current solution is added to the RMP. 

Once the violated constraints are added to the model, the column generation procedure should be restarted. Adding more constraints to the model modifies the solution space and new columns with negative reduced cost can be found. The overall procedure of column generation and separation is summarized in Algorithm \ref{alg:colgenandsep}.

\begin{algorithm}
\caption{Column generation and Separation pseudo-code} \label{alg:colgenandsep}
\begin{algorithmic}[1]
\Procedure{colGenAndSep(fixedNodes)}{}
    \State $x = \{\}$ \Comment{start with empty solution}
    \State $PP \gets fixedNodes$ \Comment{fix nodes in graphs}
    \Repeat
        \Repeat
            \State $x \gets solve(RMP)$
            \ForAll{lines}
                \State $\lambda \gets solve(PP(line))$ \Comment{generate a new column}
                \If{$\hat{c}(\lambda) < 0$} 
                \State $RMP \gets \lambda$     \Comment{add column with negative reduced cost}
                \EndIf
            \EndFor
        \Until{no more columns with negative reduced cost}
        \State $RMP \gets violatedConstraints(x)$
    \Until{no more violated constraints}
    \State \textbf{return} $x$
\EndProcedure
\end{algorithmic}
\end{algorithm}

\subsection{Dive heuristic} \label{sec:dive}
\begin{figure}[th]
    \centering
    \includegraphics[width=0.8\linewidth]{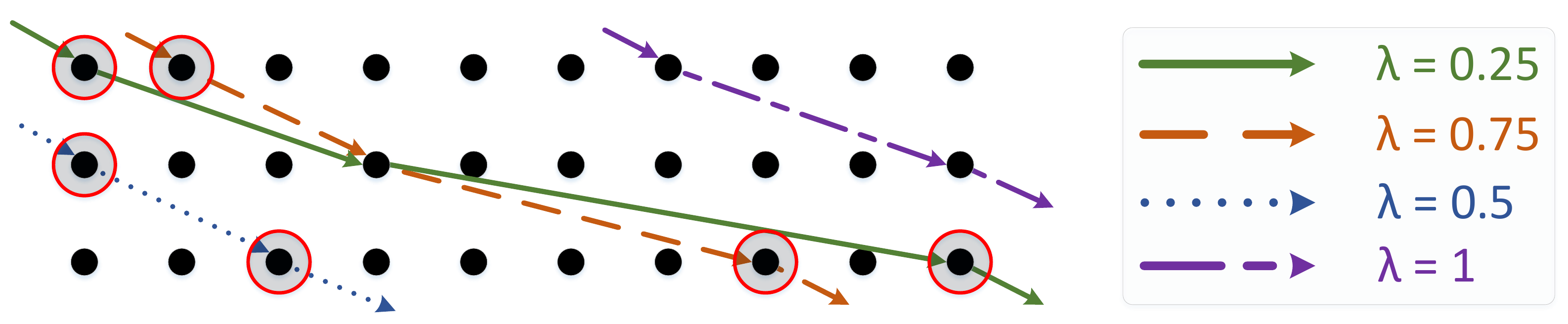}
    \caption{Fragment of a graph containing paths from a fractional solution where the nodes in the red circles are fractionally used.}
    \label{fig:FractNodes}
\end{figure}
The optimal solution for the MP can be fractional. In order to find an integer solution, a \textit{dive} heuristic method is applied.
The solution $\lambda_q$ values are added to each of the graph nodes affected by that column. This measures the "usage" of each node and, if the solution is fractional, this means that some of the graph nodes are fractionally used (see Figure \ref{fig:FractNodes}). 
The \textit{dive} heuristic selects one of the fractionally used nodes and enforces to be part of the final solution, meaning that the final integer solution must contain that node.
In order to do that, the shortest path problem is divided into two smaller and simpler ones where the chosen node works as the destination vertex in one of them and as the origin vertex in the other one.
Apart from fixing the node, all the previously generated columns from the same graph that do not include the node need to be removed from the RMP. 
Once the heuristic step is concluded, the column generation should be started again as new promising columns may be generated. 
One advantage of the dive heuristic is that it can lead faster to an integer feasible solution. A disadvantage of this method is that some branches of the tree are left unexplored and forcing the integrality of specific nodes that were fractional can lead to an infeasible final solution.
 This method only considers valid the solutions where all the trains of each line are scheduled. Therefore, in this study, if any column of the initial dummy set is part of a solution, the solution is considered infeasible. However, the initial set of dummy columns could potentially be used to allow solutions with fewer scheduled trains. 

Most of the times, there are multiple fractionally used nodes in the solution and a criterion to select one is needed. 
In this study, we have opted for choosing any fractional node at random.

The procedure is summarized in Algorithm \ref{alg:diveheuristic}.

\begin{algorithm}
\caption{Dive heuristic pseudo-code} \label{alg:diveheuristic}
\begin{algorithmic}[1]
\Procedure{diveHeuristic()}{}
    \State $[fixedNodes] = \{\}$ \Comment{initialize empty list}
    \Repeat
        \State $x \gets colGenAndSep(fixedNodes)$ \Comment{generate LP solution}
        \If{$x$ is fractional}
            \State $[fixedNodes] \gets newNode$ \Comment{fix a new node}
        \EndIf
    \Until{$x$ is integer or infeasible}
    \State \textbf{return} $x$ 
\EndProcedure
\end{algorithmic}
\end{algorithm}

\subsection{Passenger routing} \label{sec:ptt}
The main objective of the model is to improve the passenger travel time (PTT). So far, the method minimizes the length of the train paths. This avoids extra additional dwelling of the trains at the stations and allows passengers traveling in the train to reach their destination fast. However, many passengers are required to transfer between trains to reach their destinations. 
Therefore, minimizing these transfer times becomes part of the overall objective of optimizing the passenger travel time. 

  Given a timetable solution to the RMP, the total passenger travel time is computed by solving (\ref{mcfp:obj})-(\ref{mcfp:pos}).

\subsection{Benders' cuts}\label{sec:benders}
After solving (\ref{mcfp:obj})-(\ref{mcfp:pos}) for computing the total passenger travel time, we can generate a Benders' optimality cut for the RMP based on the dual values of the solution. A solution to the original RMP, fractional or not, always allows to route all the passengers and therefore, feasibility cuts are not generated.
We define variable $z_k \geq 0$ for each commodity $k\in K$.
Let $\pi^1_k,\pi^2_k \in \mathbb{R}$ be the dual variables related to constraints (\ref{mcfp:orig}) and (\ref{mcfp:dest}) respectively. Additionally, we denote $\pi^3_{kv} \in \mathbb{R}$ to the dual variable of constraint (\ref{mcfp:flow}) and $\pi^4_{ka} \leq 0$ to the dual variable of constraint (\ref{mcfp:allow}). 
The arising optimality cut for each commodity $k$ can be formulated as follows:
\begin{equation}\label{eq:benders}
    z_k \geq \pi^1_k - \pi^2_k + \sum_{a \in \bar{A}_r} \pi^4_{ka}x_a
\end{equation}
These cuts are added to the RMP and the objective function is updated to account for the $z_k$ variables as follows:
\begin{equation}
    \min \sum_{q \in Q} c_q \lambda_q + \alpha\sum_{k \in K}OD_k z_k
\end{equation}
where $\alpha \in \mathbb{R}^+$ is a parameter that defines the weight of the passenger travel time in the objective function.  
The reader may wonder why we do not just minimize $\sum_{k\in K}OD_k z_k$ since this is the true objective considered in this paper (minimizing passenger travel time). The reason is that adding Benders' cuts slows the processing of each node in the dive-tree significantly, and therefore it is not possible to add all the Benders' cuts that actually are violated if we want the overall algorithm to finish within reasonable time (we use a one hour time limit in the computational results). Therefore, we only add a subset of the violated Benders cuts and the $z_k$ variables are only an approximation of the true passenger travel time. This implies that it is beneficial to keep the path length component of the objective function as it guides the search towards solutions that also are attractive from a passenger travel time point of view.
The cuts can be added by separation in the same way as the constraints mentioned in Section \ref{sec:separation}.

As a timetable solution is given as input, most of the arcs of the routing graph are not enabled, meaning that $ x_a = 0$ in constraint (\ref{mcfp:allow}). Looking at the objective function of the dual problem, which corresponds to the right-hand side of equation (\ref{eq:benders}), the dual values of these arcs can fluctuate unrealistically as their value does not longer have an effect on the objective value of the dual problem. This can have a potential negative effect on the quality of the generated cuts.
In order to avoid this, for all arcs where no flow should be allowed, we enable an infinitesimally small capacity $\epsilon$. If $ x_a$ is not strictly zero for any arc in the graph anymore, the dual variables $\pi^4_{ka}$ are expected to be more realistic while still obtaining a near-optimal flow. This $\epsilon$-capacity method is based on the \textit{Kelley+} approach suggested by \cite{fischetti2017a}.

\subsection{Large Neighborhood Search} \label{sec:lns}

The main objective of the algorithm is to minimize the PTT. Therefore, every time a solution is computed, its PTT is compared with the best one found so far and updated if the new one is better. 
The process 
is framed in a Large Neighborhood Search (LNS) proposed by \cite{shaw1998a} where the current solution is iteratively transformed into a different one. The transformation occurs by partially destroying the current solution and repairing it again. Our LNS is inspired by the work of \cite{ropke2006} and the whole process is summarized in Algorithm \ref{alg:algorithm}.

\begin{algorithm}
\caption{Large neighborhood search pseudo-code} \label{alg:algorithm}
\begin{algorithmic}[1]
\Procedure{LNS()}{}
    \Repeat
        \State $x \gets diveHeuristic()$ \Comment{generate an initial solution}
    \Until{$x$ is feasible}
    \State $x^b = x$ 
    \Repeat
         \State $x^t \gets repair(destroy(x))$ \Comment{generate a new solution}
         \If{$x^t$ is feasible}
            \State $c(x^t) \gets routing(x^t)$ \Comment{compute PTT based on the routing of passengers}
            \If{$c(x^t) < c(x^b)$} \Comment{compare passenger travel time}
                    \State $x^b = x^t$
                    \State $x = x^t$ \Comment{only accept improving solutions}
            \EndIf
        \EndIf
    \Until{time limit}
    \State \textbf{return} $x^b$ 
\EndProcedure
\end{algorithmic}
\end{algorithm}

The repair method is the already mentioned \textit{dive-and-cut-and-price} whereas
the 
the destroy method selects randomly $\rho$ graph paths from the solution and removes them. This method
is inspired by the ones implemented by \cite{barrena2014a}.
Furthermore, only solutions improving the PTT are accepted, adding relevance to the passengers' routes.

Two versions of the LNS method are implemented: (1) An LNS method without Benders' cuts and, (2) an LNS method with Beders cuts. In the latter, these cuts are added in the separation procedure, meaning that in line 14 of Algorithm \ref{alg:colgenandsep}, we compute equation (\ref{eq:benders}) together with (\ref{cons::SPcOT})-(\ref{cons::SPSib}). The potential number of violated Benders' can be very large. This can result in not being able to solve the root node within the algorithm time limit. In order to avoid this, a internal time limit is set to stop generating Benders' cuts. 

Finally, in order to analyze the quality of the solution, this is compared with a lower bound solution. The lower bound (LB) value for the total path lengths is computed as the LP solution value at the root node 
in the initial dive heuristic (line 3 in Algorithm \ref{alg:algorithm}).
The lower bound for the PTT is computed given a solution where all the trains operate at the minimum running and dwell times (i.e. shortest train paths) and passengers are able to transfer between any pair of lines at the minimum transfer time.

\subsection{Random iterative method}
An additional method to the LNS is proposed for comparison.
The \textit{dive-and-cut-and-price} procedure is repeated iteratively where each iteration is independent from the previous one. Since the randomness is introduced in the branching process, the root node is solved once and used as the re-start point at a new iteration.
The entire method is summarized in Algorithm \ref{alg:iterative}.

\begin{algorithm}
\caption{Random iterative method} \label{alg:iterative}
\begin{algorithmic}[1]
\Procedure{RandomIterative()}{}
    \State $x = \{\}$ \Comment{Initialize empty solution}
    \State $x^b = \{\}$ \Comment{Initialize best solution}
    \State $x^r \gets solveRootNode(x)$ \Comment{solve root node}
    \Repeat
         \State $x^t \gets diveHeuristic(x^r)$ \Comment{apply dive heuristic}
         \If{$x^t$ is feasible}
            \State $c(x^t) \gets MCFP(x^t)$ \Comment{compute PTT based on the routing of passengers}
            \If{$c(x^t) < c(x^b)$} \Comment{compare passenger travel time}
                    \State $x^b = x^t$
            \EndIf
        \EndIf
    \Until{time limit}
    \State \textbf{return} $x^b$ 
\EndProcedure
\end{algorithmic}
\end{algorithm}

\section{Case study}\label{sec:CaseStudy}

\begin{figure}[ht]
    \centering
    \includegraphics[width=0.7\textwidth]{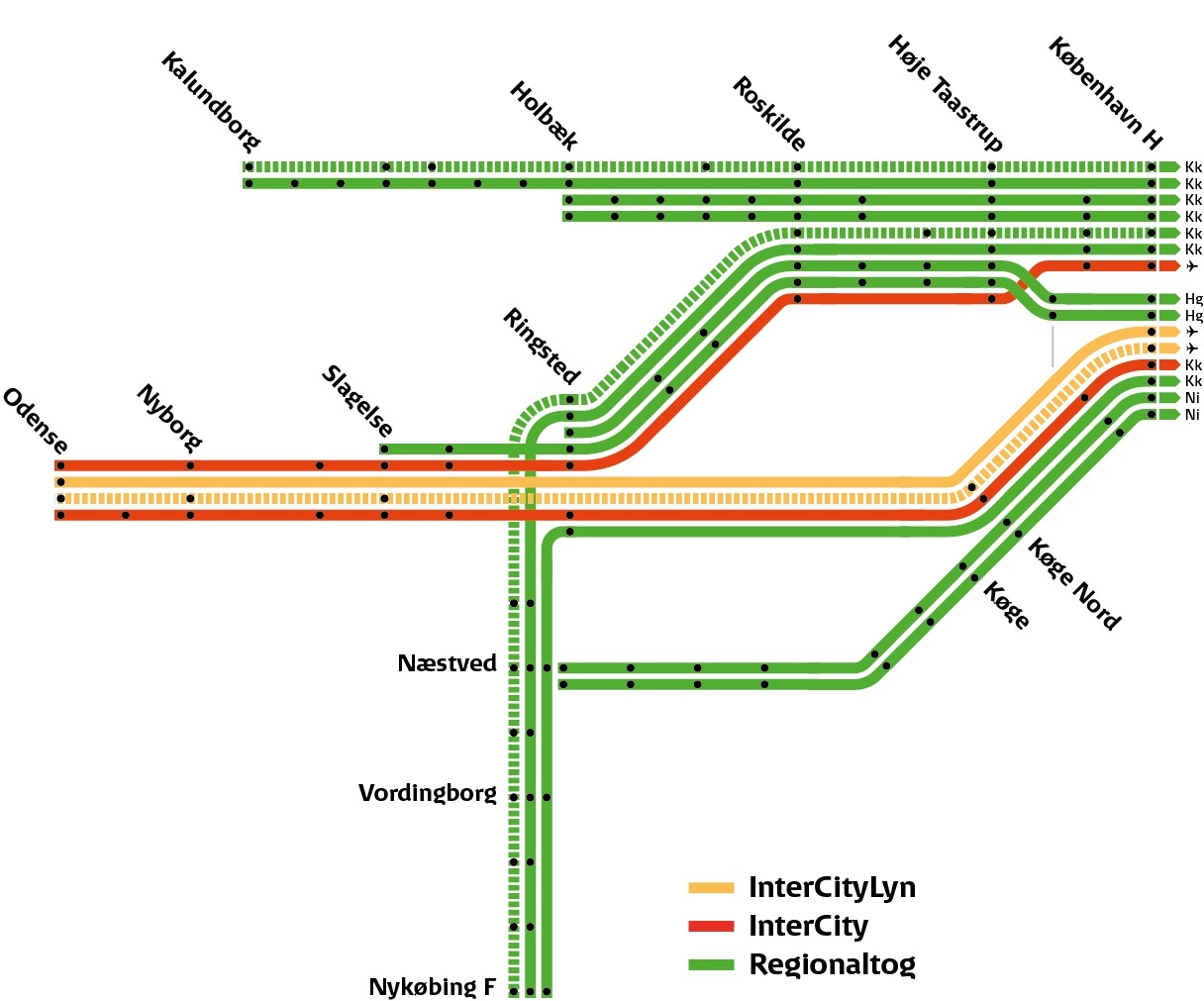}
    \caption[Network considered in the case study \citep{dsb2018}]{Network considered in the case study. Each line represents a frequency of one train per hour and direction and the dashed lines represent trains only running during rush hours \citep{dsb2018}}
    \label{fig:K19crop}
\end{figure}

The case studied here covers the Regional, Intercity and IntercityLyn (high-speed) network of Zealand, Denmark as seen in Figure \ref{fig:K19crop}.
More specifically, the scope covers a one hour period during morning rush hour.
This means that more lines run towards Copenhagen.
Once, a timetable for this period is obtained, it can be rolled out for the rest of the day by removing or adding rush hour lines.

The network is formed by 15 lines, covering 43 passenger stations. 3 of the lines only run during rush hour, which makes a total of 27 trains per hour to schedule. This translates in 12 Symmetric Line graphs, as two identical lines are handled as one line with a frequency of two trains per hour.

The number of tracks and the direction of trains running along them vary along each corridor. Three different types of track segments between stations are present in this network. A \textit{single-track} segment, where trains can circulate in both directions but there can only be one train on the segment at a time. A \textit{double-track} segment, where two tracks connect two stations allowing trains to travel in both directions (one track per direction) and a \textit{quadruple-track} segment, formed by four tracks between two consecutive stations and trains can travel in both directions (two tracks per direction). The quadruple-track segments allow two trains going in the same direction to overtake each other along the segment.

In the network considered, there are two main single-track segments: the segment between Holb\ae k and Kalundborg and the segment connecting K\o ge Nord and N\ae stved along the southern corridor. 
The rest of the network is connected by double-track segments with the exception of the segments between H\o je Taastrup and Roskilde that are formed by quadruple-tracks. 

The following input data has been provided by DSB, a danish TOC:
    
    \textbf{Minimum running time:} This parameter states the minimum required time for a train to travel between two specific stations. 
    This time interval is usually depending on the rolling stock type and the speed limits on the track segment. 
    A value is given for every track segment connecting two consecutive stations in each line and direction.
    
    \textbf{Minimum dwelling time:} This parameter states the minimum required time for a train to dwell at a specific station. 
    This time is usually the time required by the passengers to board and leave the train. 
    A value is given for every station visited by each line and each direction
    (i.e. between 30 seconds and 2 minutes).
    
    \textbf{Sibling lines:} As mentioned in Section \ref{sec:freqCons}, there are specific pairs of lines that have similar or identical routes which are required by DSB to be as separated as possible in the timeline.
    There are three pairs of these lines considered in this case study. For example, the two lines reaching Kalundborg.
    
    \textbf{Minimum headway between trains:} 
    In this case study, three minimum headway values are given: 1) Minimum headway between two consecutive departing trains in the same track segment and direction, 2) minimum headway between two consecutive arriving trains in the same track and direction and 3) minimum headway between two consecutive trains arriving from single-tracks in opposite directions.
    
    \textbf{Single-platform stations:} Some stations along the single-track segments have only one platform meaning that the station can only host one train at a time and a crossing between two trains is not allowed. 
    It is assumed that, for the rest of stations in the network, any train arriving from an adjacent track segment has an available arriving platform. 
    
    \textbf{Origin-Destination matrix:} This matrix defines the number of passengers per hour traveling between each pair of stations. It does not consider passengers from stations outside the network (i.e. people entering the network from Germany or cities in Jutland).
    There is a total of 1806 pairs.
    
\begin{figure}[ht]
    \centering
    \includegraphics[width=0.8\textwidth]{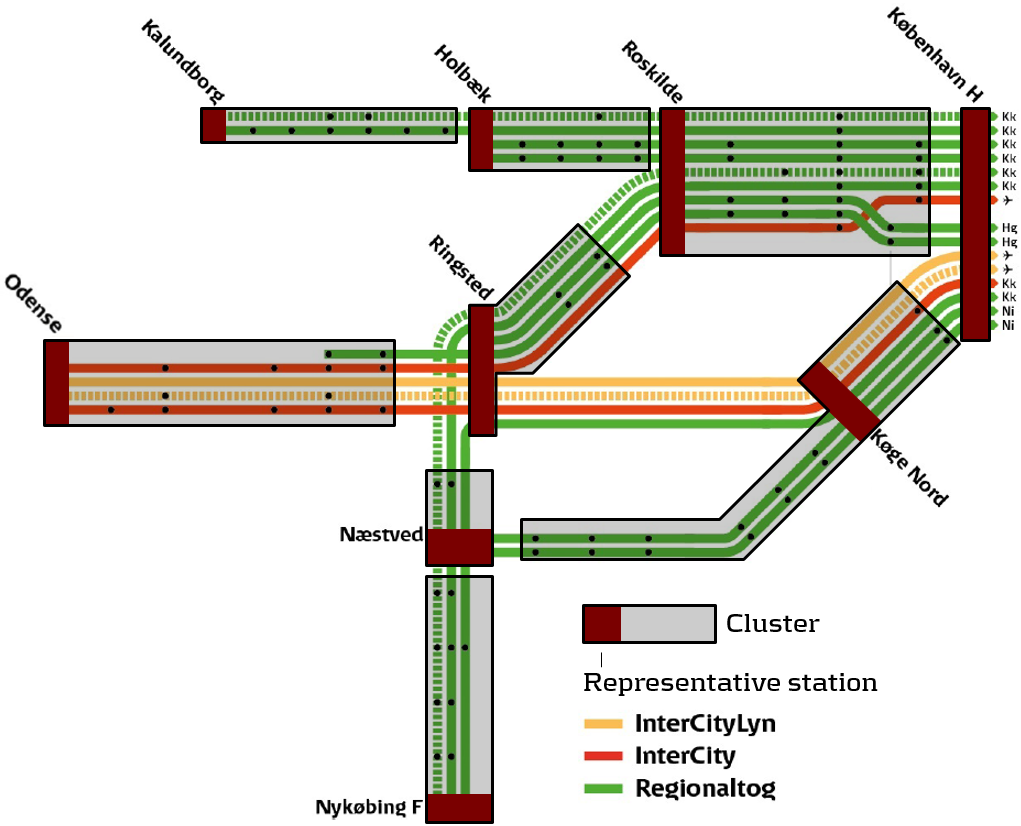}
    \caption{Network considered in the case study where the stations are divided into clusters, and the one marked in red is the representative station. Based on \citep{dsb2018}.}
    \label{fig:K19clusters}
    \end{figure}

    \textbf{Station Clusters:} A reduced version of the origin-destination pairs is proposed by defining a set of representative stations in the network, and clustering the neighboring ones.
    As shown in Figure \ref{fig:K19clusters}, these stations correspond to the end-of line stations and stations where the track segments branch in different directions (i.e. Roskilde and K\o ge Nord). The purpose of this network setup is only to reduce the number of commodities $K$ considered when generating Bender's cuts while still capturing the routing decisions of most passengers. The evaluation of the total passenger travel time in the rest of the solution method is done using the entire network.
    
    \textbf{Minimum transfer time:} In order for passengers to transfer between trains at a station, a minimum transfer time of 5 minutes is defined as a rule of thumb, meaning that if the time difference between the arrival of one train and the departure of another is lower, the transfer time corresponds to the time interval plus $|T|$.

The authors refer to \cite{martin2018a} for further details on the case study.

\subsection{Instances}
A number of instances are created based on the data from DSB. By changing the following four parameters, a total of
21
instances are obtained.

    $\boldsymbol{HW_k}$: Minimum headway between consecutive arrivals and departures at K\o benhavn H. This station is seen as one of the most congested stations in the network where all lines stop at and, therefore, the headway at this station becomes interesting to analyze individually. This parameter measures in minutes the minimum interval between consecutive arrivals or departures at K\o benhavn H in the same track segment.
    
    $\boldsymbol{HW_n}$: Minimum headway between consecutive arrivals and departures at any station in the network. This parameter measures in minutes the minimum interval between consecutive train arrivals or departures at each track segment and station in the network.
    
    $\boldsymbol{HW_s}$: Minimum headway between consecutive departures of
    sibling 
    trains in the same direction from common stations. The pair of
    sibling 
    lines may have slightly different stopping patterns or running and dwell times. This makes impossible to separate both train paths exactly half an hour during their entire trip. Therefore, a lower bound is needed that should be respected in any station. In this case, a minimum headway is defined for the consecutive departures from each station.
    
     $\boldsymbol{\kappa}$: maximum symmetry gap in $\pm$ minutes between departure and arrival of trains in opposite direction belonging to the same line. 

\subsection{Computational results}
The model has been entirely written in \textit{Julia} language \citep{bezanson2017julia}, modelled using \textit{JuMP} \citep{lubin2015computing} and using \textit{CPLEX v.
12.9} as the solver. It has been tested in an Intel Xeon Processor X5550 (quad-core, 2.66 GHz) using one thread.
Due to the large amount of parameter setting combinations, a base case is defined with the minimum values of each parameter (except for $\kappa$). Then, each parameter is tested independently keeping the others fixed. The parameter values for the base case are shown in Table \ref{tab:baseCaseParam}. All instances are tested with a maximum dwell time of 3 minutes at each station.
A parameter tuning has been conducted to determine the degree of destruction ($\rho$) of the destroy method which has been set to 5. 

\begin{table}[th]
\centering
\caption{Base case parameter setting}
\label{tab:baseCaseParam}
\begin{tabular}{cccc}
\textbf{\begin{tabular}[c]{@{}c@{}}$HW_k$\\ (min)\end{tabular}} & \textbf{\begin{tabular}[c]{@{}c@{}}$HW_n$\\ (min)\end{tabular}} & \textbf{\begin{tabular}[c]{@{}c@{}}$HW_s$\\ (min)\end{tabular}} &
\textbf{\begin{tabular}[c]{@{}c@{}}$\kappa$\\ ($\pm$min)\end{tabular}}
\\ \hline
3 & 3 & 15 & 
1.5 
\end{tabular}
\end{table}

\subsubsection{Impact of the pricing problem}

In order to measure the benefits of the new graph formulation. A variant of the method (from now on referred as \textit{Train-graph model}) is tested where the graphs only generate the train paths of a line in one direction and the symmetry is ensured by adding the respective constraints in the RMP. Due to the poor performance of the Train-graph model, only a comparison of the root node calculation is shown in Table \ref{tab:trainRN}.
For the given network, the Symmetric Line graph is able to provide a stronger lower bound in significantly less time and fewer iterations. Actually, the lower bound of the Train-graph model corresponds to the sum of minimum running and dwell times of the trains to be scheduled (i.e. constant term). 
\begin{table}[th]
\centering
\caption{Root node results of the Symmetric Line graph model and the Train graph one.}
\label{tab:trainRN}
\begin{tabular}{cccc}
\textbf{Model} & \textbf{Obj. value (min)} & \textbf{CG Iters} &
\textbf{Time (s)} \\ \hline
Train graph & 1981 & 39 &  27 \\
Symmetric Line graph & 1998.5 & 7 &  3 \\
\end{tabular}
\end{table}

In addition, the dive heuristic based on the Train graph model is not able to find a feasible solution within the 1 hour limit.
We believe that these results show that the graph formulation is an important part of the proposed solution method.

\subsubsection{Instance results}

The three solution methods presented are run 10 times for each scenario and the average values calculated. The time limit for each algorithm run is set to 1 hour and the internal time limit to stop adding Benders' cuts is set to 10 \% of the algorithm time limit (i.e. 6 minutes).
The value of $\alpha$ is set to the inverse of the number of passengers travelling within the time period.  

\begin{figure}[th]
    \centering
    \includegraphics[width=\linewidth]{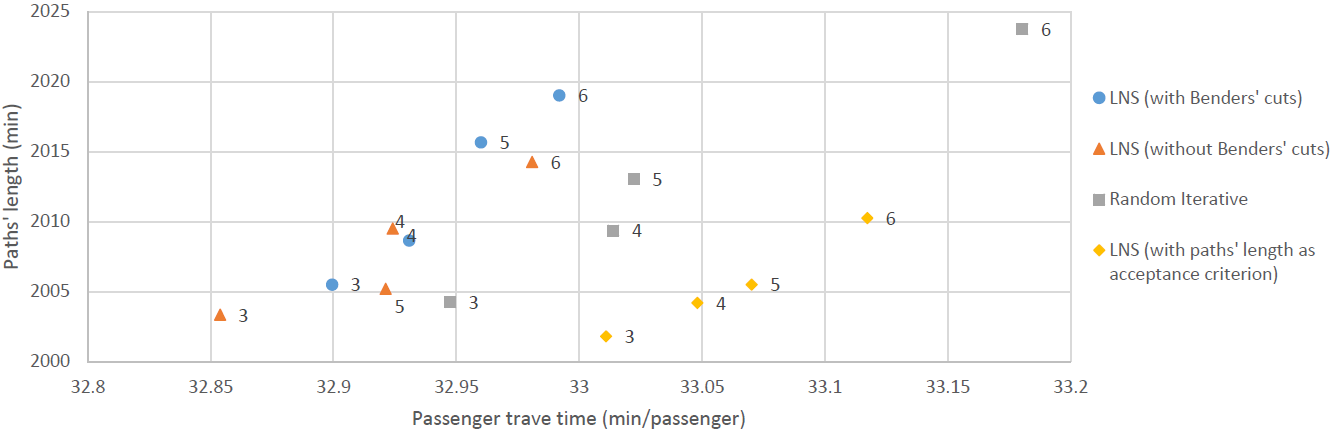}
    \caption{Average solution values instances with different values of $HW_k$ in minutes.}
    \label{fig:HWk}
\end{figure}

\begin{figure}[th]
    \centering
    \includegraphics[width=\linewidth]{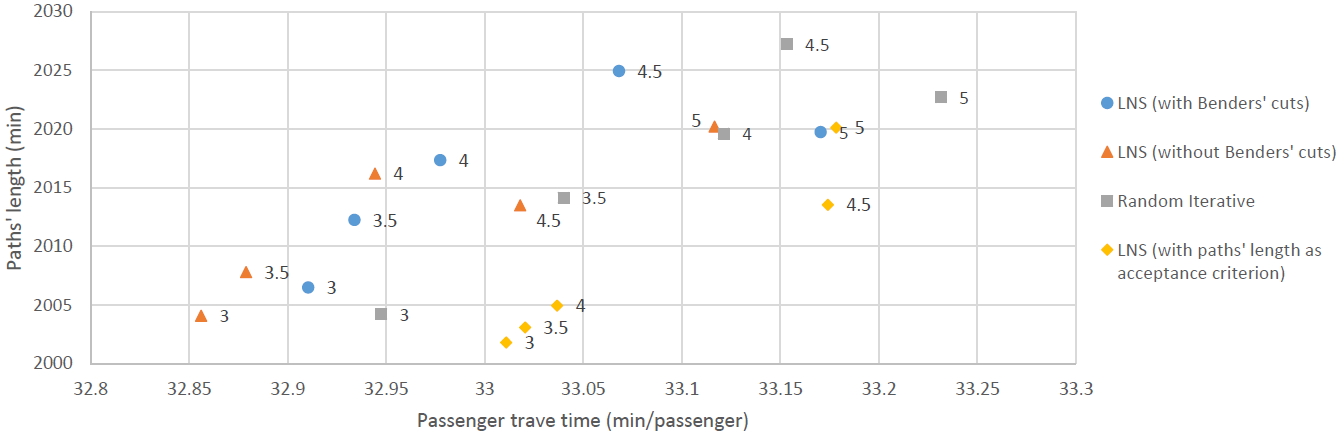}
    \caption{Average solution values instances with different values of $HW_n$ in minutes.}
    \label{fig:HWn}
\end{figure}

\begin{figure}[th]
    \centering
    \includegraphics[width=\linewidth]{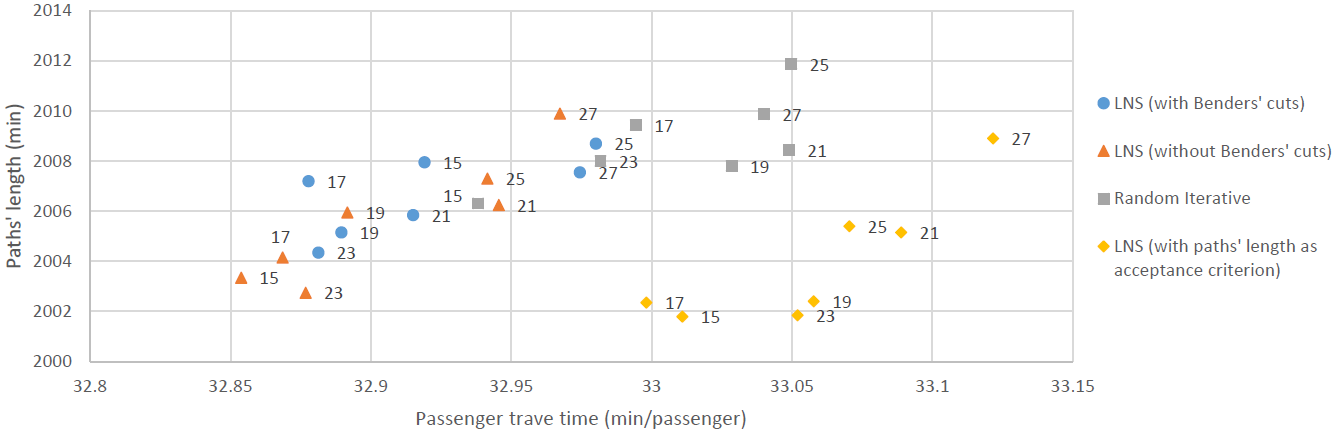}
    \caption{Average solution values instances with different values of $HW_s$ in minutes.}
    \label{fig:HWs}
\end{figure}

\begin{figure}[th]
    \centering
    \includegraphics[width=\linewidth]{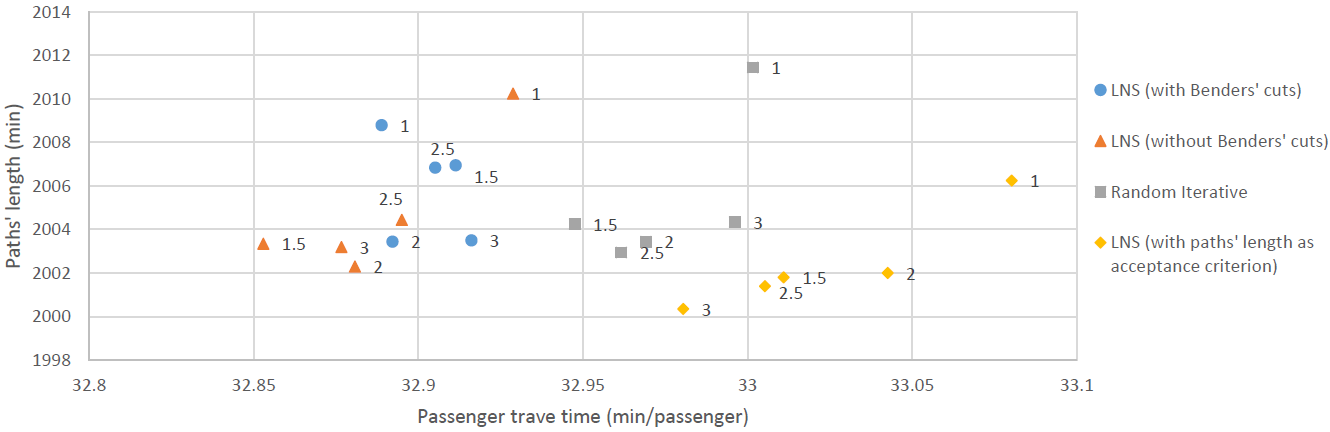}
    \caption{Average solution values instances with different values of $\kappa$ in $\pm$ minutes.}
    \label{fig:MaxG}
\end{figure}

Tables \ref{tab:hwK}-\ref{tab:maxG} show the results for each of the scenarios created by parameters $HW_k$, $HW_n$, $HW_s$ and $\kappa$ respectively.
The first column indicates the solution method and the second one the parameter value of the scenario. The third and fourth columns display the best and average solution values of PTT respectively found across the 10 runs which are compared to the lower bound defined at the end of Section \ref{sec:lns}.
The fifth and sixth columns indicate the average sum of path lengths (PL) relative to the best integer solutions with and without considering the fixed term compared to the lower bound defined at the end of Section \ref{sec:lns}.
The seventh column displays the number of algorithm iterations or equivalent repetitions of lines 7-14 in Algorithm \ref{alg:algorithm} done per 1h run. The next three columns indicate the internal average iterations per algorithm iteration. First, the number of dive heuristic iterations which can be interpreted as the number of branches performed (i.e. nodes fixed). Next, the number of times the current LP solution is checked for violated constraints and, finally, the number of column generation iterations. The eleventh and twelfth columns shows the average number of columns and additional rows (\ref{cons::SPcOT})-(\ref{cons::SPSib}) added per algorithm iteration respectively. 
The thirteenth column indicates the number of Benders' cuts added in total. 
The next three columns indicate the proportional amount of time spent solving the RMP, PP and in the separation procedure respectively in relation to the total amount of time spent finding a solution. Last, the feasibility rate is stated that displays the proportion of algorithm iterations that result in a feasible integer solution. 
The average solution values are displayed in Figures \ref{fig:HWk}-\ref{fig:MaxG} which also include results of a variant of the LNS method without Benders' cuts that only accepts solutions that improve the paths' length. More detailed results about this method variant can be found in  Table \ref{tab:SolValsPLaccept} in the Appendix.

The LNS-based methods that include passenger travel time in the acceptance criterion show a better performance in all instances. A main reason lies on the amount of iterations each method is able to perform. This suggests that partially destroying the solution is effective and enables exploring multiple neighborhoods.
Table \ref{tab:avgAll} shows the average solution quality, over all instances considered, in terms of passenger travel time and paths' length for the 4 variants of the solution method. The table shows that
the addition of Benders' cuts results in a similar but marginally worse overall solution quality than the LNS method that did not use the Benders cuts.
Adding these cuts increases the complexity of the RMP leading to fewer algorithm iterations
and this causes the two methods to end with a similar solution quality.

All the methods find near optimal results both in PTT and path lengths in a reasonable amount of time for most of the scenarios.
From Figures \ref{fig:HWk}-\ref{fig:MaxG} a correlation between the length of the paths and the passenger travel time can be inferred. This is a realistic assumption since 95.8 \% of all passengers in the network can reach their destination boarding a single train.
However, optimizing the length of the train paths does not necessarily result in a shorter passenger travel time. This is deduced from the results of the LNS method with paths' length as acceptance criterion. In general, the solutions have indeed shorter train paths but the total passenger travel time is worse than for other methods. 
This shows that the simple integration of passenger travel time objective into the LNS method, through the acceptance criterion, is important in order to reach high quality solutions.

Intuitively, the parameter with the highest impact in PTT variation is the headway at the entire network, followed by the one at Copenhagen's central station. 
The similar performance for most of the values of $HW_s$ indicates that this headway parameter has a very low impact in the solution space. The variability in the solution given by the randomness of the method allows, in cases like this where the instances are very similar, to have slightly better results even if the parameter value is more restrictive. Ideally this should not happen and we believe that a longer running time or more algorithm runs per instance would smooth the trend.
Moreover, little variation in PTT is shown for the different values of maximum symmetry gap. 
This indicates a trade-off between the maximum gap allowed and the iterations the algorithm is able to perform within the time limit. A higher value of $\kappa$, expands the solution space but fewer algorithm iterations hinder the exploration of the neighborhood efficiently. On the other hand, if $\kappa$ is too tight, the solution space becomes highly restricted and, regardless of the number of iterations, the solution quality decreases.
It should be noted that the instances with the lowest values of the headway parameters correspond to the same instance. Different randomized seeds have been used in all cases and therefore, the results are not identical.

In terms of speed, it can be seen that the problem becomes harder to solve when increasing the parameter values. In particular, for high 
$HW_n$ values, the LP becomes very hard to solve.
Likewise, a higher value of $\kappa$, increases the complexity of the graph formulation and that is reflected in the time spent solving the pricing problems.
Nevertheless, all methods are able to find solutions for $HW_k=6$ minutes which is the maximum possible as 10 trains arrive per hour in K\o benhavn H through the same corridor. Also, solutions are found for values up to $HW_n=5$ minutes and higher values were not further tested as they do not seem realistic for the network studied. Moreover, the algorithm finds solutions for $HW_s=27$ minutes which seems to be the maximum allowed due to the differences in running times of the pairs of
sibling  lines.
To put the solution values into perspective, we can compare them to the manual timetable planned by DSB ($PTT=33.21, PL=2049.5)$. It can be noticed the presented methods produce better results for all instances. However, the manual timetable considers additional operational aspects such as rolling stock assignment and track crossings and therefore, we cannot see it as a fair comparison.

\begin{landscape}
\begin{table}[]
\centering
\caption{Average performance of the algorithms for different values of $HW_k$.}
\label{tab:hwK}
\resizebox{1.6\textwidth}{!}{%
\begin{tabu}{>{\footnotesize}c|rrrrrrrrrrrrrrrr}
\rowfont{\footnotesize}
\textbf{Method} & \multicolumn{1}{c}{\textbf{\begin{tabular}[c]{@{}c@{}}$HW_k$ \\ (min)\end{tabular}}} & \multicolumn{1}{c}{\textbf{\begin{tabular}[c]{@{}c@{}}Best PTT \\ gap (\%)\end{tabular}}} & \multicolumn{1}{c}{\textbf{\begin{tabular}[c]{@{}c@{}}PTT \\ gap (\%)\end{tabular}}} & \multicolumn{1}{c}{\textbf{\begin{tabular}[c]{@{}c@{}}PL \\ gap (\%)\end{tabular}}} & \multicolumn{1}{c}{\textbf{\begin{tabular}[c]{@{}c@{}}Var. PL \\ gap (\%)\end{tabular}}} & \multicolumn{1}{c}{\textbf{\begin{tabular}[c]{@{}c@{}}Alg \\ It.\end{tabular}}} & \multicolumn{1}{c}{\textbf{\begin{tabular}[c]{@{}c@{}}Avg \\ Dive It.\end{tabular}}} & \multicolumn{1}{c}{\textbf{\begin{tabular}[c]{@{}c@{}}Avg \\ Sep It.\end{tabular}}} & \multicolumn{1}{c}{\textbf{\begin{tabular}[c]{@{}c@{}}Avg \\ CG It.\end{tabular}}} & \multicolumn{1}{c}{\textbf{\begin{tabular}[c]{@{}c@{}}Avg \\ Columns\end{tabular}}} & \multicolumn{1}{c}{\textbf{\begin{tabular}[c]{@{}c@{}}Avg \\ +rows (\%)\end{tabular}}} & \multicolumn{1}{c}{\textbf{\begin{tabular}[c]{@{}c@{}}Benders' \\ cuts\end{tabular}}} & \multicolumn{1}{c}{\textbf{\begin{tabular}[c]{@{}c@{}}Avg RMP \\ time (\%)\end{tabular}}} & \multicolumn{1}{c}{\textbf{\begin{tabular}[c]{@{}c@{}}Avg PP \\ time (\%)\end{tabular}}} & \multicolumn{1}{c}{\textbf{\begin{tabular}[c]{@{}c@{}}Avg Sep \\ time (\%)\end{tabular}}} & \multicolumn{1}{c}{\textbf{\begin{tabular}[c]{@{}c@{}}Feasibility \\ rate (\%)\end{tabular}}} \\ \hline
\multirow{4}{*}{\textbf{\begin{tabular}[c]{@{}c@{}}LNS \\ (with \\ Benders'\\ cuts)\end{tabular}}} & 3 & 2.54 & 2.75 & 0.35 & 40.0 & 80.6 & 2.4 & 6.0 & 125.0 & 299.5 & 1.41 & 124 & 38 & 34 & 13 & 91 \\
 & 4 & 2.61 & 2.85 & 0.51 & 58.0 & 77.4 & 2.2 & 5.6 & 126.5 & 345.3 & 1.04 & 156 & 40 & 31 & 15 & 90 \\
 & 5 & 2.72 & 2.94 & 0.86 & 98.0 & 83.3 & 2.0 & 5.0 & 107.6 & 315.8 & 0.79 & 182 & 39 & 32 & 14 & 90 \\
 & 6 & 2.71 & 3.04 & 1.03 & 117.1 & 85.3 & 1.7 & 4.1 & 94.5 & 313.5 & 0.56 & 177 & 42 & 29 & 13 & 89 \\ \hline
\multirow{4}{*}{\textbf{\begin{tabular}[c]{@{}c@{}}LNS\\ (without\\ Benders'\\ cuts)\end{tabular}}} & 3 & 2.26 & 2.61 & 0.24 & 27.7 & 143.6 & 0.6 & 3.9 & 98.7 & 238.4 & 1.03 & 0 & 40 & 28 & 4 & 94 \\
 & 4 & 2.37 & 2.83 & 0.55 & 62.9 & 134.1 & 0.7 & 3.8 & 115.2 & 294.7 & 1.08 & 0 & 43 & 30 & 4 & 88 \\
 & 5 & 2.56 & 2.82 & 0.34 & 38.3 & 169.3 & 0.5 & 3.1 & 74.1 & 218.1 & 0.59 & 0 & 37 & 27 & 4 & 93 \\
 & 6 & 2.71 & 3.00 & 0.79 & 90.0 & 164.2 & 0.6 & 2.9 & 79.2 & 245.8 & 0.44 & 0 & 40 & 27 & 3 & 92 \\ \hline
\multirow{4}{*}{\textbf{\begin{tabular}[c]{@{}c@{}}Random \\ Iterative\end{tabular}}} & 3 & 2.64 & 2.90 & 0.29 & 32.9 & 27.6 & 6.8 & 22.5 & 227.3 & 773.0 & 11.85 & 0 & 62 & 31 & 5 & 44 \\
 & 4 & 2.82 & 3.11 & 0.54 & 62.0 & 19.0 & 7.0 & 23.2 & 272.3 & 999.1 & 9.82 & 0 & 72 & 24 & 3 & 45 \\
 & 5 & 2.83 & 3.13 & 0.73 & 83.1 & 16.7 & 7.1 & 23.1 & 316.7 & 1251.6 & 8.97 & 0 & 85 & 13 & 1 & 41 \\
 & 6 & 2.82 & 3.63 & 1.26 & 144.3 & 8.7 & 6.1 & 18.1 & 351.7 & 1688.6 & 4.93 & 0 & 87 & 12 & 1 & 36
\end{tabu}
}
\end{table}

\begin{table}[]
\centering
\caption{Average performance of the algorithms for different values of $HW_n$.}
\label{tab:hwN}
\resizebox{1.6\textwidth}{!}{%
\begin{tabu}{>{\footnotesize}c|rrrrrrrrrrrrrrrr}
\rowfont{\footnotesize}
\textbf{Method} & \multicolumn{1}{c}{\textbf{\begin{tabular}[c]{@{}c@{}}$HW_n$ \\ (min)\end{tabular}}} & \multicolumn{1}{c}{\textbf{\begin{tabular}[c]{@{}c@{}}Best PTT \\ gap (\%)\end{tabular}}} & \multicolumn{1}{c}{\textbf{\begin{tabular}[c]{@{}c@{}}PTT \\ gap (\%)\end{tabular}}} & \multicolumn{1}{c}{\textbf{\begin{tabular}[c]{@{}c@{}}PL \\ gap (\%)\end{tabular}}} & \multicolumn{1}{c}{\textbf{\begin{tabular}[c]{@{}c@{}}Var. PL \\ gap (\%)\end{tabular}}} & \multicolumn{1}{c}{\textbf{\begin{tabular}[c]{@{}c@{}}Alg \\ It.\end{tabular}}} & \multicolumn{1}{c}{\textbf{\begin{tabular}[c]{@{}c@{}}Avg \\ Dive It.\end{tabular}}} & \multicolumn{1}{c}{\textbf{\begin{tabular}[c]{@{}c@{}}Avg \\ Sep It.\end{tabular}}} & \multicolumn{1}{c}{\textbf{\begin{tabular}[c]{@{}c@{}}Avg \\ CG It.\end{tabular}}} & \multicolumn{1}{c}{\textbf{\begin{tabular}[c]{@{}c@{}}Avg \\ Columns\end{tabular}}} & \multicolumn{1}{c}{\textbf{\begin{tabular}[c]{@{}c@{}}Avg \\ +rows (\%)\end{tabular}}} & \multicolumn{1}{c}{\textbf{\begin{tabular}[c]{@{}c@{}}Benders' \\ cuts\end{tabular}}} & \multicolumn{1}{c}{\textbf{\begin{tabular}[c]{@{}c@{}}Avg RMP \\ time (\%)\end{tabular}}} & \multicolumn{1}{c}{\textbf{\begin{tabular}[c]{@{}c@{}}Avg PP \\ time (\%)\end{tabular}}} & \multicolumn{1}{c}{\textbf{\begin{tabular}[c]{@{}c@{}}Avg Sep \\ time (\%)\end{tabular}}} & \multicolumn{1}{c}{\textbf{\begin{tabular}[c]{@{}c@{}}Feasibility \\ rate (\%)\end{tabular}}} \\ \hline
\multirow{5}{*}{\textbf{\begin{tabular}[c]{@{}c@{}}LNS \\ (with \\ Benders'\\ cuts)\end{tabular}}} & 3 & 2.54 & 2.78 & 0.40 & 45.7 & 74.9 & 2.2 & 5.9 & 125.5 & 304.5 & 1.44 & 119 & 38 & 35 & 14 & 90 \\
\multicolumn{1}{c|}{} & 3.5 & 2.51 & 2.86 & 0.69 & 78.6 & 72.3 & 2.1 & 5.6 & 120.8 & 320.0 & 1.29 & 118 & 39 & 34 & 14 & 92 \\
\multicolumn{1}{c|}{} & 4 & 2.69 & 2.99 & 0.94 & 107.7 & 55.8 & 2.2 & 6.1 & 147.2 & 464.5 & 1.85 & 156 & 46 & 31 & 14 & 88 \\
\multicolumn{1}{c|}{} & 4.5 & 2.96 & 3.27 & 1.32 & 151.1 & 48.5 & 2.3 & 7.5 & 220.3 & 851.8 & 3.51 & 131 & 66 & 21 & 11 & 76 \\
\multicolumn{1}{c|}{} & 5 & 2.84 & 3.59 & 1.06 & 121.4 & 25.3 & 3.1 & 10.6 & 393.3 & 1620.5 & 7.13 & 137 & 84 & 7 & 8 & 78 \\ \hline
\multirow{5}{*}{\textbf{\begin{tabular}[c]{@{}c@{}}LNS\\ (without\\ Benders'\\ cuts)\end{tabular}}} & 3 & 2.36 & 2.61 & 0.28 & 32.0 & 139.4 & 0.6 & 3.8 & 100.7 & 242.4 & 1.04 & 0 & 41 & 28 & 4 & 95 \\
\multicolumn{1}{c|}{} & 3.5 & 2.43 & 2.68 & 0.47 & 53.1 & 128.5 & 0.5 & 3.5 & 104.5 & 276.5 & 1.03 & 0 & 52 & 25 & 3 & 93 \\
\multicolumn{1}{c|}{} & 4 & 2.46 & 2.89 & 0.89 & 101.1 & 123.3 & 0.5 & 3.2 & 107.7 & 311.3 & 1.01 & 0 & 45 & 30 & 4 & 91 \\
\multicolumn{1}{c|}{} & 4.5 & 2.76 & 3.12 & 0.75 & 85.7 & 96.7 & 1.0 & 5.7 & 149.7 & 513.5 & 3.73 & 0 & 70 & 22 & 2 & 83 \\
\multicolumn{1}{c|}{} & 5 & 2.82 & 3.43 & 1.09 & 124.0 & 66.9 & 1.7 & 9.6 & 323.4 & 1273.7 & 6.91 & 0 & 93 & 6 & 0 & 88 \\ \hline
\multirow{5}{*}{\textbf{\begin{tabular}[c]{@{}c@{}}Random \\ Iterative\end{tabular}}} & 3 & 2.64 & 2.90 & 0.29 & 32.9 & 28.0 & 6.9 & 22.6 & 227.9 & 776.3 & 11.86 & 0 & 61 & 32 & 5 & 43 \\
\multicolumn{1}{c|}{} & 3.5 & 2.73 & 3.19 & 0.78 & 89.4 & 15.9 & 6.5 & 22.6 & 290.4 & 1084.8 & 14.10 & 0 & 87 & 11 & 1 & 41 \\
\multicolumn{1}{c|}{} & 4 & 2.83 & 3.44 & 1.06 & 120.6 & 7.5 & 6.1 & 23.9 & 388.8 & 1576.7 & 16.64 & 0 & 92 & 7 & 1 & 41 \\
\multicolumn{1}{c|}{} & 4.5 & 3.13 & 3.54 & 1.44 & 164.3 & 5.1 & 5.4 & 22.1 & 484.0 & 1975.8 & 18.91 & 0 & 94 & 5 & 0 & 35 \\
\multicolumn{1}{c|}{} & 5 & 3.47 & 3.78 & 1.21 & 138.3 & 3.8 & 4.8 & 18.8 & 492.7 & 2247.1 & 17.78 & 0 & 97 & 3 & 0 & 35
\end{tabu}
}
\end{table}

\begin{table}[]
\centering
\caption{Average performance of the algorithms for different values of $HW_s$.}
\label{tab:hwS}
\resizebox{1.6\textwidth}{!}{%
\begin{tabu}{>{\footnotesize}c|rrrrrrrrrrrrrrrr}
\rowfont{\footnotesize}
\textbf{Method} & \multicolumn{1}{c}{\textbf{\begin{tabular}[c]{@{}c@{}}$HW_s$ \\ (min)\end{tabular}}} & \multicolumn{1}{c}{\textbf{\begin{tabular}[c]{@{}c@{}}Best PTT \\ gap (\%)\end{tabular}}} & \multicolumn{1}{c}{\textbf{\begin{tabular}[c]{@{}c@{}}PTT \\ gap (\%)\end{tabular}}} & \multicolumn{1}{c}{\textbf{\begin{tabular}[c]{@{}c@{}}PL \\ gap (\%)\end{tabular}}} & \multicolumn{1}{c}{\textbf{\begin{tabular}[c]{@{}c@{}}Var. PL \\ gap (\%)\end{tabular}}} & \multicolumn{1}{c}{\textbf{\begin{tabular}[c]{@{}c@{}}Alg \\ It.\end{tabular}}} & \multicolumn{1}{c}{\textbf{\begin{tabular}[c]{@{}c@{}}Avg \\ Dive It.\end{tabular}}} & \multicolumn{1}{c}{\textbf{\begin{tabular}[c]{@{}c@{}}Avg \\ Sep It.\end{tabular}}} & \multicolumn{1}{c}{\textbf{\begin{tabular}[c]{@{}c@{}}Avg \\ CG It.\end{tabular}}} & \multicolumn{1}{c}{\textbf{\begin{tabular}[c]{@{}c@{}}Avg \\ Columns\end{tabular}}} & \multicolumn{1}{c}{\textbf{\begin{tabular}[c]{@{}c@{}}Avg \\ +rows (\%)\end{tabular}}} & \multicolumn{1}{c}{\textbf{\begin{tabular}[c]{@{}c@{}}Benders' \\ cuts\end{tabular}}} & \multicolumn{1}{c}{\textbf{\begin{tabular}[c]{@{}c@{}}Avg RMP \\ time (\%)\end{tabular}}} & \multicolumn{1}{c}{\textbf{\begin{tabular}[c]{@{}c@{}}Avg PP \\ time (\%)\end{tabular}}} & \multicolumn{1}{c}{\textbf{\begin{tabular}[c]{@{}c@{}}Avg Sep \\ time (\%)\end{tabular}}} & \multicolumn{1}{c}{\textbf{\begin{tabular}[c]{@{}c@{}}Feasibility \\ rate (\%)\end{tabular}}} \\ \hline
\multirow{7}{*}{\textbf{\begin{tabular}[c]{@{}c@{}}LNS \\ (with \\ Benders'\\ cuts)\end{tabular}}} & 15 & 2.60 & 2.81 & 0.47 & 54.0 & 81.9 & 2.2 & 5.8 & 130.0 & 312.2 & 1.38 & 129 & 40 & 32 & 13 & 92 \\
\multicolumn{1}{c|}{} & 17 & 2.51 & 2.68 & 0.44 & 49.7 & 75.6 & 2.4 & 6.1 & 133.9 & 329.3 & 1.62 & 142 & 39 & 33 & 14 & 89 \\
\multicolumn{1}{c|}{} & 19 & 2.39 & 2.72 & 0.33 & 38.0 & 84.0 & 2.4 & 5.8 & 119.7 & 298.2 & 1.64 & 123 & 37 & 33 & 14 & 92 \\
\multicolumn{1}{c|}{} & 21 & 2.37 & 2.80 & 0.37 & 42.0 & 75.2 & 2.1 & 5.5 & 131.2 & 352.8 & 1.66 & 135 & 42 & 31 & 13 & 91 \\
\multicolumn{1}{c|}{} & 23 & 2.50 & 2.69 & 0.29 & 33.4 & 81.3 & 2.4 & 5.7 & 113.2 & 295.1 & 1.90 & 150 & 37 & 32 & 15 & 91 \\
\multicolumn{1}{c|}{} & 25 & 2.77 & 3.00 & 0.51 & 58.3 & 86.5 & 1.9 & 5.5 & 113.2 & 281.1 & 2.12 & 115 & 36 & 34 & 15 & 89 \\
\multicolumn{1}{c|}{} & 27 & 2.53 & 2.98 & 0.45 & 51.7 & 69.9 & 1.9 & 6.0 & 131.6 & 361.4 & 3.30 & 112 & 37 & 36 & 16 & 78 \\ \hline
\multirow{7}{*}{\textbf{\begin{tabular}[c]{@{}c@{}}LNS\\ (without\\ Benders'\\ cuts)\end{tabular}}} & 15 & 2.26 & 2.61 & 0.24 & 27.7 & 142.2 & 0.6 & 3.9 & 98.7 & 238.3 & 1.03 & 0 & 41 & 28 & 4 & 94 \\
\multicolumn{1}{c|}{} & 17 & 2.45 & 2.65 & 0.28 & 32.3 & 139.3 & 0.7 & 3.9 & 102.6 & 245.4 & 1.22 & 0 & 41 & 29 & 5 & 92 \\
\multicolumn{1}{c|}{} & 19 & 2.49 & 2.72 & 0.37 & 42.6 & 136.9 & 0.6 & 3.7 & 104.9 & 246.2 & 1.27 & 0 & 41 & 29 & 5 & 90 \\
\multicolumn{1}{c|}{} & 21 & 2.49 & 2.89 & 0.39 & 44.3 & 126.1 & 0.8 & 4.5 & 110.5 & 278.3 & 2.00 & 0 & 41 & 32 & 6 & 88 \\
\multicolumn{1}{c|}{} & 23 & 2.37 & 2.68 & 0.21 & 24.3 & 155.3 & 0.4 & 3.3 & 85.2 & 210.5 & 1.38 & 0 & 37 & 27 & 7 & 92 \\
\multicolumn{1}{c|}{} & 25 & 2.62 & 2.88 & 0.44 & 50.3 & 132.3 & 0.5 & 3.9 & 106.8 & 251.3 & 1.99 & 0 & 39 & 29 & 8 & 91 \\
\multicolumn{1}{c|}{} & 27 & 2.71 & 2.96 & 0.57 & 65.1 & 133.3 & 0.5 & 3.6 & 94.8 & 236.2 & 2.29 & 0 & 38 & 29 & 9 & 88 \\ \hline
\multirow{7}{*}{\textbf{\begin{tabular}[c]{@{}c@{}}Random \\ Iterative\end{tabular}}} & 15 & 2.64 & 2.87 & 0.39 & 44.6 & 28.0 & 6.9 & 22.5 & 222.6 & 763.0 & 11.92 & 0 & 61 & 31 & 5 & 43 \\
\multicolumn{1}{c|}{} & 17 & 2.86 & 3.04 & 0.55 & 62.5 & 24.3 & 7.0 & 23.4 & 229.5 & 835.1 & 13.01 & 0 & 84 & 13 & 2 & 53 \\
\multicolumn{1}{c|}{} & 19 & 2.86 & 3.15 & 0.47 & 53.2 & 24.6 & 6.7 & 21.4 & 215.9 & 777.6 & 12.76 & 0 & 79 & 18 & 3 & 31 \\
\multicolumn{1}{c|}{} & 21 & 2.94 & 3.21 & 0.50 & 56.8 & 19.3 & 7.1 & 23.6 & 288.3 & 989.1 & 14.52 & 0 & 87 & 11 & 1 & 35 \\
\multicolumn{1}{c|}{} & 23 & 2.75 & 3.01 & 0.48 & 54.3 & 23.1 & 6.1 & 19.8 & 202.9 & 798.0 & 13.80 & 0 & 84 & 13 & 2 & 49 \\
\multicolumn{1}{c|}{} & 25 & 2.95 & 3.22 & 0.67 & 76.4 & 16.3 & 6.3 & 20.7 & 278.7 & 1090.8 & 15.49 & 0 & 91 & 8 & 1 & 48 \\
\multicolumn{1}{c|}{} & 27 & 2.79 & 3.19 & 0.57 & 65.0 & 36.3 & 4.2 & 13.2 & 143.3 & 662.3 & 14.10 & 0 & 80 & 16 & 3 & 13
\end{tabu}
}
\end{table}

\begin{table}[]
\centering
\caption{Average performance of the algorithms for different values of $\kappa$.}
\label{tab:maxG}
\resizebox{1.6\textwidth}{!}{%
\begin{tabu}{>{\footnotesize}c|rrrrrrrrrrrrrrrr}
\rowfont{\footnotesize}
\textbf{Method} & \multicolumn{1}{c}{\textbf{\begin{tabular}[c]{@{}c@{}}$\kappa$ \\ ($\pm$ min)\end{tabular}}} & \multicolumn{1}{c}{\textbf{\begin{tabular}[c]{@{}c@{}}Best PTT \\ gap (\%)\end{tabular}}} & \multicolumn{1}{c}{\textbf{\begin{tabular}[c]{@{}c@{}}PTT \\ gap (\%)\end{tabular}}} & \multicolumn{1}{c}{\textbf{\begin{tabular}[c]{@{}c@{}}PL \\ gap (\%)\end{tabular}}} & \multicolumn{1}{c}{\textbf{\begin{tabular}[c]{@{}c@{}}Var. PL \\ gap (\%)\end{tabular}}} & \multicolumn{1}{c}{\textbf{\begin{tabular}[c]{@{}c@{}}Alg \\ It.\end{tabular}}} & \multicolumn{1}{c}{\textbf{\begin{tabular}[c]{@{}c@{}}Avg \\ Dive It.\end{tabular}}} & \multicolumn{1}{c}{\textbf{\begin{tabular}[c]{@{}c@{}}Avg \\ Sep It.\end{tabular}}} & \multicolumn{1}{c}{\textbf{\begin{tabular}[c]{@{}c@{}}Avg \\ CG It.\end{tabular}}} & \multicolumn{1}{c}{\textbf{\begin{tabular}[c]{@{}c@{}}Avg \\ Columns\end{tabular}}} & \multicolumn{1}{c}{\textbf{\begin{tabular}[c]{@{}c@{}}Avg \\ +rows (\%)\end{tabular}}} & \multicolumn{1}{c}{\textbf{\begin{tabular}[c]{@{}c@{}}Benders' \\ cuts\end{tabular}}} & \multicolumn{1}{c}{\textbf{\begin{tabular}[c]{@{}c@{}}Avg RMP \\ time (\%)\end{tabular}}} & \multicolumn{1}{c}{\textbf{\begin{tabular}[c]{@{}c@{}}Avg PP \\ time (\%)\end{tabular}}} & \multicolumn{1}{c}{\textbf{\begin{tabular}[c]{@{}c@{}}Avg Sep \\ time (\%)\end{tabular}}} & \multicolumn{1}{c}{\textbf{\begin{tabular}[c]{@{}c@{}}Feasibility \\ rate (\%)\end{tabular}}} \\ \hline
\multirow{5}{*}{\textbf{\begin{tabular}[c]{@{}c@{}}LNS \\ (with \\ Benders'\\ cuts)\end{tabular}}} & 1 & 2.43 & 2.71 & 0.39 & 39.0 & 93.6 & 2.2 & 5.7 & 116.3 & 283.1 & 1.33 & 162 & 41 & 26 & 16 & 90 \\
\multicolumn{1}{c|}{} & 1.5 & 2.52 & 2.79 & 0.42 & 48.3 & 76.9 & 2.3 & 5.9 & 127.5 & 315.3 & 1.36 & 129 & 38 & 34 & 13 & 92 \\
\multicolumn{1}{c|}{} & 2 & 2.30 & 2.73 & 0.30 & 36.1 & 72.6 & 2.1 & 5.4 & 121.5 & 311.5 & 1.17 & 117 & 34 & 39 & 13 & 92 \\
\multicolumn{1}{c|}{} & 2.5 & 2.40 & 2.77 & 0.47 & 56.7 & 57.6 & 2.6 & 6.3 & 148.5 & 359.3 & 1.47 & 127 & 34 & 43 & 13 & 90 \\
\multicolumn{1}{c|}{} & 3 & 2.53 & 2.80 & 0.30 & 36.4 & 50.3 & 2.9 & 7.0 & 154.0 & 402.7 & 1.55 & 124 & 36 & 44 & 12 & 92 \\ \hline
\multirow{5}{*}{\textbf{\begin{tabular}[c]{@{}c@{}}LNS\\ (without\\ Benders'\\ cuts)\end{tabular}}}  & 1 & 2.57 & 2.84 & 0.46 & 46.3 & 162.1 & 0.6 & 3.9 & 99.4 & 237.3 & 1.15 & 0 & 44 & 22 & 5 & 93 \\
\multicolumn{1}{c|}{} & 1.5 & 2.26 & 2.60 & 0.24 & 27.7 & 146.4 & 0.6 & 3.9 & 98.8 & 238.4 & 1.03 & 0 & 41 & 28 & 4 & 94 \\
\multicolumn{1}{c|}{} & 2 & 2.42 & 2.69 & 0.24 & 29.1 & 129.6 & 0.7 & 3.7 & 102.3 & 248.6 & 1.07 & 0 & 38 & 34 & 4 & 92 \\
\multicolumn{1}{c|}{} & 2.5 & 2.46 & 2.73 & 0.35 & 42.1 & 100.7 & 0.8 & 4.6 & 124.5 & 299.0 & 1.43 & 0 & 36 & 42 & 4 & 91 \\
\multicolumn{1}{c|}{} & 3 & 2.30 & 2.68 & 0.29 & 34.5 & 100.9 & 0.6 & 3.9 & 109.9 & 268.3 & 1.14 & 0 & 33 & 45 & 4 & 93 \\ \hline
\multirow{5}{*}{\textbf{\begin{tabular}[c]{@{}c@{}}Random \\ Iterative\end{tabular}}} & 1 & 2.93 & 3.07 & 0.52 & 52.2 & 34.4 & 7.0 & 23.2 & 207.4 & 746.1 & 11.57 & 0 & 66 & 25 & 6 & 49 \\
\multicolumn{1}{c|}{} & 1.5 & 2.64 & 2.90 & 0.29 & 32.9 & 27.5 & 6.9 & 22.6 & 229.9 & 779.7 & 11.87 & 0 & 62 & 31 & 4 & 43 \\
\multicolumn{1}{c|}{} & 2 & 2.79 & 2.97 & 0.30 & 36.1 & 19.2 & 7.0 & 23.8 & 256.8 & 908.0 & 11.97 & 0 & 73 & 23 & 2 & 53 \\
\multicolumn{1}{c|}{} & 2.5 & 2.71 & 2.94 & 0.27 & 33.0 & 18.7 & 7.2 & 25.2 & 298.1 & 993.2 & 10.37 & 0 & 77 & 20 & 2 & 66 \\
\multicolumn{1}{c|}{} & 3 & 2.76 & 3.05 & 0.34 & 41.5 & 14.5 & 6.8 & 23.3 & 289.4 & 1000.9 & 13.70 & 0 & 70 & 27 & 2 & 50
\end{tabu}
}
\end{table}

\end{landscape}

\begin{table}[]
\centering
\caption{Average solution quality over all instances}
\label{tab:avgAll}
\begin{tabular}{c|c|c|c|c}
 & \textbf{\begin{tabular}[c]{@{}c@{}}LNS (with \\ Benders' cuts)\end{tabular}} & \textbf{\begin{tabular}[c]{@{}c@{}}LNS (without \\ Benders' cuts)\end{tabular}} & \textbf{\begin{tabular}[c]{@{}c@{}}Random \\ iterative\end{tabular}} & \textbf{\begin{tabular}[c]{@{}c@{}}LNS (with paths' length \\ as acceptance criterion)\end{tabular}} \\ \hline
\textbf{Passenger travel time} & 2.88 \% & 2.81 \% & 3.15 \% & 3.24 \% \\ \hline
\textbf{Paths' length} & 0.57 \% & 0.45 \% & 0.62 \% & 0.33 \%
\end{tabular}
\end{table}


\subsubsection{Effect of Benders' cuts}

The addition of a limited amount of Benders' cuts within the presented method does not have a significant impact. 
Additional tests were carried out where the time limit is extended so that more Benders' cuts can be generated. Nevertheless, the solution method does not provide better solutions. In fact, the quality decreases. It is observed that both generating more cuts or using a larger value of $\alpha$, increases the fractionality of the solutions and the complexity of the RMP. Furthermore, solutions with a high number of columns with small coefficients hinder the dive heuristic procedure.

Alternatively, in order to assess the potential impact of the Bender cuts, we solve the root node adding all violated Benders' cuts and compare the solution value with the manual lower bound mentioned at the end of Section \ref{sec:lns}. 
For this experiment, the objective function is modified such that only passenger travel time is considered.
This is tested on the instance with base parameter values considering the network with station clusters. The results are summarized in Table \ref{tab:LBBenders} where the first column shows the optimality gap from the precomputed manual lower bound to the best known integer solution. The second column shows the optimality gap from the root node solution solved. The number of cuts and computational time is shown in the last two columns respectively.
The results indicate that the Benders' cuts are able to provide a stronger lower bound. 
\begin{table}[]
\centering
\caption{Lower bound (LB) comparison for the base case instance with cluster stations.}
\label{tab:LBBenders}
\begin{tabular}{c|ccc}
\textbf{Manual lower bound} & \textbf{LB with Benders' cuts} & \textbf{Benders' cuts} & \textbf{Time (s)} \\ \hline
0.72 \% & 0.61 \% & 926 & 5885
\end{tabular}
\end{table}
It should be noted that solving the entire branch-and-bound tree adding all the necessary violated Benders' cuts guarantees converging to the optimal solution.
This suggest that addition of Benders' cuts may be more interesting in an exact method for the integrated model compared to in a heuristic as proposed here. It is clear, however, that such an exact method only could solve much smaller instances compared to those considered in this paper.

\section{Conclusion}\label{sec:conclusion}

As future work, it would be interesting to study how to decrease
the time for solving the LP relaxation of the master problem as this can become quite excessive for the more constrained instances. One may attempt to 1) leave the headway constraints out of the initial formulation and add the violated ones by separation or 2) attempt to stabilize the dual variables in the column generation algorithm (see e.g. \cite{du1999stabilized} or \cite{oukil2007stabilized}).

Computational results showed that the addition of Benders' cuts in the LNS did not improve the performance of the heuristic even though the cuts allow us to fully integrate the passenger travel time objective in the mathematical model that is the foundation of the LNS heuristic. It is possible that further work could change this conclusion. If the time for solving the passenger flow sub-problem could be reduced and convergence of the Benders' algorithm could be improved, such that fewer Benders' cuts are needed then the approach may be more competitive. Improved convergence of the Benders' algorithm could perhaps be achieved using generic speed-up techniques as those suggested in \cite{magnanti1981a},\cite{papadakos2008a} and \cite{fischetti2017a}.

When looking at the number of columns needed per iteration, it is also interesting to look from which lines the columns mainly come from. 
In average, 93\% of all the columns generated
belong to lines using the quadruple-track segment.
Allowing two routes for the trains doubles the number of possible columns that can be generated. It should also be noted that 32 \% of the total amount of columns belong to the two lines running until Kalundborg. This is related to the fact that at the single-track segment of this corridor, is the only place where a crossing between trains of different lines can occur. In order to cross, one of the trains needs to dwell for three minutes in one of the stations resulting in a poor path length. As the crossing constraints are added by separation, this results in a larger amount of columns generated.

The model is able to route the passengers realistically. This is analyzed using graphical tools such as the one shown in Figure \ref{fig:ChordKH} which shows the passenger flow between trains at K\o benhavn H for an example solution. Nevertheless, a more realistic routing of the passengers in the most congested areas can help to have a complete perspective of the trips of the passengers and the occupancy of the trains. This can be further improved by taking train capacity into account \citep{rezanova2015line} or achieving a more accurate estimation of the passenger demand. 

Although the fixed running times between stations simulate realistic cases to a large extent, considering variable running times at the track segments can increase significantly the solution space. However, the complexity of the model would increase accordingly. Also, considering different types of headway along the network allows a better utilization of the track capacity as more trains can be scheduled per corridor \citep{LiuHan2017}.

Different graphical tools have been used to analyze the potential additional conflicts of a timetable such as the one in Figure \ref{fig:NWtimetable} which shows an example graphic timetable for the north-western corridor between K\o benhavn H and Kalundborg.
Routing the trains at a more detailed level at some stations can allow having completely conflict-free solutions in the network. Currently, feasibility issues may arise from the model due to track-crossing conflicts at some stations where corridors join. This can be solved by adding additional graph nodes to model the track junctions. Likewise, turnaround times for trains at the end of stations can be enforced by removing the conflicting arcs in the graph. This can potentially lead to a better utilization of the rolling stock.

In conclusion, this paper aims at optimizing the railway timetable generation process from a passenger perspective. 
Solution methods have been implemented to solve the network for Regional and InterCity trains in Zealand.
The methods are based on a graph formulation that takes advantage of the symmetric timetabling strategy and the assumed fixed train running times between stations. As a result, all the required train paths for a line in a cycle time of one hour can be computed by a single \textit{shortest path}. Furthermore, the algorithms rely mainly on both column generation and constraint separation techniques. This, combined with
Benders' cuts that guide the routing of the passengers
results in an algorithm for railway timetabling 
that optimizes passenger travel time.
The methods have been shown to find good solutions to the network in a relatively fast time. The minimum headway can be easily increased along the network, achieving more robust timetables, without a significant detriment in time or solution quality.

Last but not least, the graph representation of the problem has the potential to easily model parts of the network in more detail such as track-crossing conflicts or platform assignment.
The methods can potentially be improved and implemented as a useful tool in the planning process of a train operating company.

\section*{Acknowledgements}

The work of Stefan Ropke was funded by the Innovation Fund Denmark under the IPTOP project, this support is gratefully acknowledged.
The authors are grateful 
to Federico Farina for his valuable comments and useful discussions. Moreover, thanks are due
to Esben Linde from DSB for having provided relevant data for the instances presented in the paper.
Last but not least, the authors are thankful to the Editor Prof. Oliveira and to five anonymous reviewers for their constructive remarks that have helped improve the manuscript.

\bibliography{sample}

\appendix

\section{}
\label{app:paper}
\begin{landscape}
\begin{table}[]
\centering
\caption{Average performance of LNS method with paths' length as acceptance criterion.}
\label{tab:SolValsPLaccept}
\resizebox{1.6\textwidth}{!}{%
\begin{tabular}{rrrrrrrrrrrrrrrrr}
\multicolumn{1}{c}{\textbf{\begin{tabular}[c]{@{}c@{}}$HW_k$ \\ (min)\end{tabular}}} & \multicolumn{1}{c}{\textbf{\begin{tabular}[c]{@{}c@{}}Best PTT \\ gap (\%)\end{tabular}}} & \multicolumn{1}{c}{\textbf{\begin{tabular}[c]{@{}c@{}}PTT \\ gap (\%)\end{tabular}}} & \multicolumn{1}{c}{\textbf{\begin{tabular}[c]{@{}c@{}}PL \\ gap (\%)\end{tabular}}} & \multicolumn{1}{c}{\textbf{\begin{tabular}[c]{@{}c@{}}Var. PL \\ gap (\%)\end{tabular}}} & \multicolumn{1}{c}{\textbf{\begin{tabular}[c]{@{}c@{}}Alg \\ It.\end{tabular}}} & \multicolumn{1}{c}{\textbf{\begin{tabular}[c]{@{}c@{}}Avg \\ Dive It.\end{tabular}}} & \multicolumn{1}{c}{\textbf{\begin{tabular}[c]{@{}c@{}}Avg \\ Sep It.\end{tabular}}} & \multicolumn{1}{c}{\textbf{\begin{tabular}[c]{@{}c@{}}Avg \\ CG It.\end{tabular}}} & \multicolumn{1}{c}{\textbf{\begin{tabular}[c]{@{}c@{}}Avg \\ Columns\end{tabular}}} & \multicolumn{1}{c}{\textbf{\begin{tabular}[c]{@{}c@{}}Avg \\ +rows (\%)\end{tabular}}} & \multicolumn{1}{c}{\textbf{\begin{tabular}[c]{@{}c@{}}Benders \\ cuts\end{tabular}}} & \multicolumn{1}{c}{\textbf{\begin{tabular}[c]{@{}c@{}}Avg RMP \\ time (\%)\end{tabular}}} & \multicolumn{1}{c}{\textbf{\begin{tabular}[c]{@{}c@{}}Avg PP \\ time (\%)\end{tabular}}} & \multicolumn{1}{c}{\textbf{\begin{tabular}[c]{@{}c@{}}Avg Sep \\ time (\%)\end{tabular}}} & \multicolumn{1}{c}{\textbf{\begin{tabular}[c]{@{}c@{}}Feasibility \\ rate (\%)\end{tabular}}} \\ \hline
3 & 2.59 & 3.10 & 0.17 & 18.86 & 143.7 & 0.6 & 3.7 & 97.7 & 242.7 & 1.01 & 0 & 40 & 29 & 4 & 93 \\
4 & 2.87 & 3.21 & 0.29 & 32.57 & 146.5 & 0.6 & 3.5 & 96.8 & 251.1 & 0.83 & 0 & 40 & 29 & 4 & 92 \\
5 & 2.84 & 3.28 & 0.35 & 40.00 & 166.5 & 0.5 & 3.1 & 76.4 & 225.8 & 0.60 & 0 & 37 & 28 & 4 & 93 \\
6 & 3.08 & 3.43 & 0.59 & 67.14 & 167.0 & 0.5 & 2.9 & 75.5 & 236.2 & 0.48 & 0 & 39 & 27 & 4 & 91 \\ \hline
\multicolumn{1}{c}{\textbf{\begin{tabular}[c]{@{}c@{}}$HW_n$ \\ (min)\end{tabular}}} & \multicolumn{1}{c}{\textbf{\begin{tabular}[c]{@{}c@{}}Best PTT \\ gap (\%)\end{tabular}}} & \multicolumn{1}{c}{\textbf{\begin{tabular}[c]{@{}c@{}}PTT \\ gap (\%)\end{tabular}}} & \multicolumn{1}{c}{\textbf{\begin{tabular}[c]{@{}c@{}}PL \\ gap (\%)\end{tabular}}} & \multicolumn{1}{c}{\textbf{\begin{tabular}[c]{@{}c@{}}Var. PL \\ gap (\%)\end{tabular}}} & \multicolumn{1}{c}{\textbf{\begin{tabular}[c]{@{}c@{}}Alg \\ It.\end{tabular}}} & \multicolumn{1}{c}{\textbf{\begin{tabular}[c]{@{}c@{}}Avg \\ Dive It.\end{tabular}}} & \multicolumn{1}{c}{\textbf{\begin{tabular}[c]{@{}c@{}}Avg \\ Sep It.\end{tabular}}} & \multicolumn{1}{c}{\textbf{\begin{tabular}[c]{@{}c@{}}Avg \\ CG It.\end{tabular}}} & \multicolumn{1}{c}{\textbf{\begin{tabular}[c]{@{}c@{}}Avg \\ Columns\end{tabular}}} & \multicolumn{1}{c}{\textbf{\begin{tabular}[c]{@{}c@{}}Avg \\ +rows (\%)\end{tabular}}} & \multicolumn{1}{c}{\textbf{\begin{tabular}[c]{@{}c@{}}Benders \\ cuts\end{tabular}}} & \multicolumn{1}{c}{\textbf{\begin{tabular}[c]{@{}c@{}}Avg RMP \\ time (\%)\end{tabular}}} & \multicolumn{1}{c}{\textbf{\begin{tabular}[c]{@{}c@{}}Avg PP \\ time (\%)\end{tabular}}} & \multicolumn{1}{c}{\textbf{\begin{tabular}[c]{@{}c@{}}Avg Sep \\ time (\%)\end{tabular}}} & \multicolumn{1}{c}{\textbf{\begin{tabular}[c]{@{}c@{}}Feasibility \\ rate (\%)\end{tabular}}} \\ \hline
3 & 2.59 & 3.10 & 0.17 & 18.9 & 144.1 & 0.6 & 3.7 & 98.0 & 243.5 & 1.01 & 0 & 40 & 29 & 4 & 93 \\
3.5 & 2.71 & 3.13 & 0.23 & 26.3 & 148.3 & 0.5 & 3.2 & 92.9 & 248.8 & 0.93 & 0 & 40 & 29 & 4 & 92 \\
4 & 2.72 & 3.18 & 0.32 & 36.9 & 143.5 & 0.4 & 2.9 & 87.3 & 253.9 & 0.80 & 0 & 40 & 29 & 4 & 94 \\
4.5 & 3.04 & 3.61 & 0.75 & 86.0 & 89.1 & 0.7 & 4.3 & 124.7 & 417.6 & 2.35 & 0 & 58 & 28 & 3 & 87 \\
5 & 3.05 & 3.62 & 1.08 & 123.5 & 51.0 & 1.9 & 9.8 & 311.7 & 1240.4 & 7.70 & 0 & 91 & 7 & 1 & 71 \\ \hline
\multicolumn{1}{c}{\textbf{\begin{tabular}[c]{@{}c@{}}$HW_s$ \\ (min)\end{tabular}}} & \multicolumn{1}{c}{\textbf{\begin{tabular}[c]{@{}c@{}}Best PTT \\ gap (\%)\end{tabular}}} & \multicolumn{1}{c}{\textbf{\begin{tabular}[c]{@{}c@{}}PTT \\ gap (\%)\end{tabular}}} & \multicolumn{1}{c}{\textbf{\begin{tabular}[c]{@{}c@{}}PL \\ gap (\%)\end{tabular}}} & \multicolumn{1}{c}{\textbf{\begin{tabular}[c]{@{}c@{}}Var. PL \\ gap (\%)\end{tabular}}} & \multicolumn{1}{c}{\textbf{\begin{tabular}[c]{@{}c@{}}Alg \\ It.\end{tabular}}} & \multicolumn{1}{c}{\textbf{\begin{tabular}[c]{@{}c@{}}Avg \\ Dive It.\end{tabular}}} & \multicolumn{1}{c}{\textbf{\begin{tabular}[c]{@{}c@{}}Avg \\ Sep It.\end{tabular}}} & \multicolumn{1}{c}{\textbf{\begin{tabular}[c]{@{}c@{}}Avg \\ CG It.\end{tabular}}} & \multicolumn{1}{c}{\textbf{\begin{tabular}[c]{@{}c@{}}Avg \\ Columns\end{tabular}}} & \multicolumn{1}{c}{\textbf{\begin{tabular}[c]{@{}c@{}}Avg \\ +rows (\%)\end{tabular}}} & \multicolumn{1}{c}{\textbf{\begin{tabular}[c]{@{}c@{}}Benders \\ cuts\end{tabular}}} & \multicolumn{1}{c}{\textbf{\begin{tabular}[c]{@{}c@{}}Avg RMP \\ time (\%)\end{tabular}}} & \multicolumn{1}{c}{\textbf{\begin{tabular}[c]{@{}c@{}}Avg PP \\ time (\%)\end{tabular}}} & \multicolumn{1}{c}{\textbf{\begin{tabular}[c]{@{}c@{}}Avg Sep \\ time (\%)\end{tabular}}} & \multicolumn{1}{c}{\textbf{\begin{tabular}[c]{@{}c@{}}Feasibility \\ rate (\%)\end{tabular}}} \\ \hline
15 & 2.59 & 3.10 & 0.17 & 18.9 & 140.8 & 0.6 & 3.7 & 97.7 & 242.8 & 1.01 & 0 & 40 & 29 & 4 & 93 \\
17 & 2.57 & 3.06 & 0.19 & 22.0 & 141.2 & 0.6 & 3.9 & 95.9 & 228.9 & 1.20 & 0 & 39 & 29 & 5 & 92 \\
19 & 2.89 & 3.24 & 0.20 & 22.3 & 137.3 & 0.5 & 3.5 & 100.6 & 247.5 & 1.37 & 0 & 40 & 29 & 5 & 92 \\
21 & 2.98 & 3.34 & 0.33 & 38.0 & 127.8 & 0.8 & 4.5 & 106.9 & 268.4 & 2.01 & 0 & 40 & 32 & 6 & 87 \\
23 & 2.75 & 3.22 & 0.17 & 19.1 & 147.8 & 0.5 & 3.3 & 89.1 & 221.3 & 1.44 & 0 & 37 & 28 & 7 & 93 \\
25 & 3.09 & 3.28 & 0.35 & 39.4 & 128.6 & 0.5 & 3.7 & 106.5 & 260.0 & 2.06 & 0 & 39 & 30 & 8 & 90 \\
27 & 2.79 & 3.44 & 0.52 & 59.4 & 124.6 & 0.6 & 3.8 & 101.1 & 251.1 & 2.49 & 0 & 38 & 30 & 9 & 88 \\ \hline
\multicolumn{1}{c}{\textbf{\begin{tabular}[c]{@{}c@{}}$\kappa$ \\ ($\pm$ min)\end{tabular}}} & \multicolumn{1}{c}{\textbf{\begin{tabular}[c]{@{}c@{}}Best PTT \\ gap (\%)\end{tabular}}} & \multicolumn{1}{c}{\textbf{\begin{tabular}[c]{@{}c@{}}PTT \\ gap (\%)\end{tabular}}} & \multicolumn{1}{c}{\textbf{\begin{tabular}[c]{@{}c@{}}PL \\ gap (\%)\end{tabular}}} & \multicolumn{1}{c}{\textbf{\begin{tabular}[c]{@{}c@{}}Var. PL \\ gap (\%)\end{tabular}}} & \multicolumn{1}{c}{\textbf{\begin{tabular}[c]{@{}c@{}}Alg \\ It.\end{tabular}}} & \multicolumn{1}{c}{\textbf{\begin{tabular}[c]{@{}c@{}}Avg \\ Dive It.\end{tabular}}} & \multicolumn{1}{c}{\textbf{\begin{tabular}[c]{@{}c@{}}Avg \\ Sep It.\end{tabular}}} & \multicolumn{1}{c}{\textbf{\begin{tabular}[c]{@{}c@{}}Avg \\ CG It.\end{tabular}}} & \multicolumn{1}{c}{\textbf{\begin{tabular}[c]{@{}c@{}}Avg \\ Columns\end{tabular}}} & \multicolumn{1}{c}{\textbf{\begin{tabular}[c]{@{}c@{}}Avg \\ +rows (\%)\end{tabular}}} & \multicolumn{1}{c}{\textbf{\begin{tabular}[c]{@{}c@{}}Benders \\ cuts\end{tabular}}} & \multicolumn{1}{c}{\textbf{\begin{tabular}[c]{@{}c@{}}Avg RMP \\ time (\%)\end{tabular}}} & \multicolumn{1}{c}{\textbf{\begin{tabular}[c]{@{}c@{}}Avg PP \\ time (\%)\end{tabular}}} & \multicolumn{1}{c}{\textbf{\begin{tabular}[c]{@{}c@{}}Avg Sep \\ time (\%)\end{tabular}}} & \multicolumn{1}{c}{\textbf{\begin{tabular}[c]{@{}c@{}}Feasibility \\ rate (\%)\end{tabular}}} \\ \hline
1 & 2.68 & 3.31 & 0.26 & 26.2 & 158.7 & 0.6 & 3.8 & 96.8 & 230.4 & 1.11 & 0 & 43 & 22 & 5 & 92 \\
1.5 & 2.59 & 3.10 & 0.17 & 18.9 & 140.0 & 0.6 & 3.7 & 98.2 & 243.6 & 1.01 & 0 & 40 & 29 & 4 & 93 \\
2 & 2.85 & 3.19 & 0.23 & 27.3 & 128.4 & 0.6 & 3.6 & 98.8 & 241.8 & 0.99 & 0 & 37 & 35 & 4 & 93 \\
2.5 & 2.75 & 3.08 & 0.20 & 23.6 & 107.4 & 0.8 & 4.3 & 107.5 & 263.5 & 1.30 & 0 & 34 & 41 & 4 & 93 \\
3 & 2.34 & 3.00 & 0.14 & 17.3 & 107.0 & 0.6 & 3.7 & 98.9 & 255.0 & 1.01 & 0 & 31 & 44 & 4 & 93
\end{tabular}
}
\end{table}
\end{landscape}
\begin{figure}[ht]
    \centering
    \includegraphics[width=\linewidth]{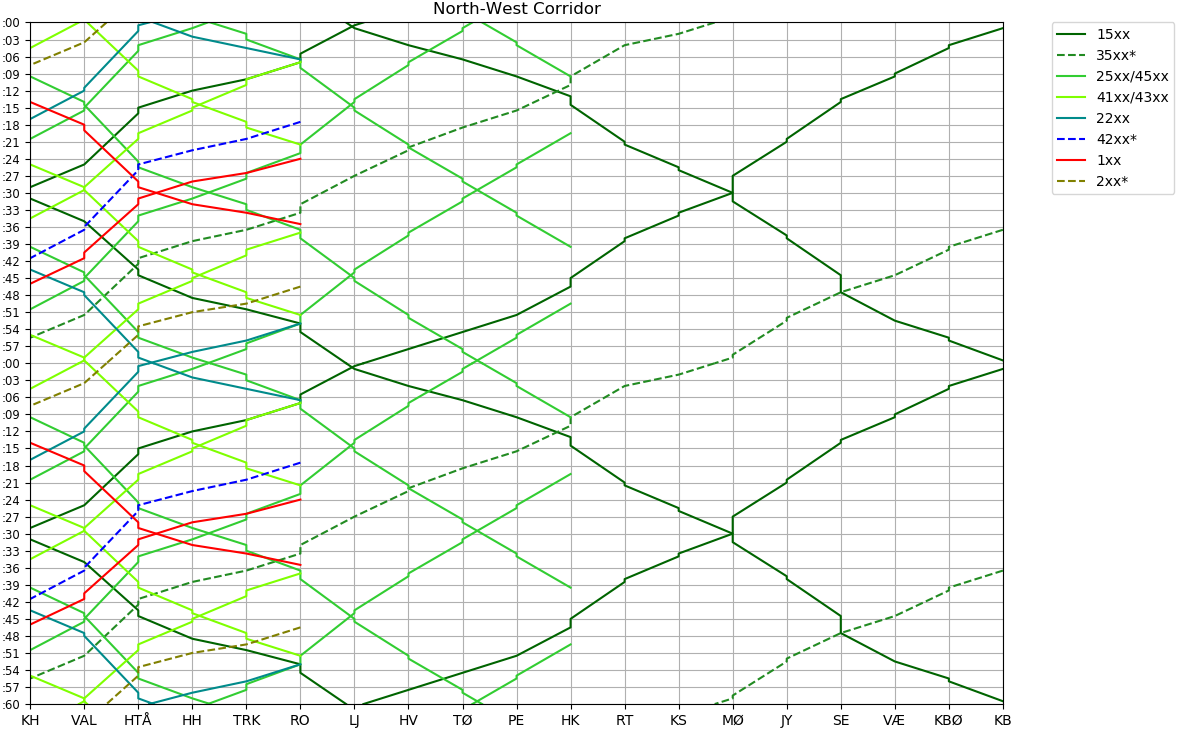}
    \caption{Timetable example for the lines running through the North-West corridor}
    \label{fig:NWtimetable}
\end{figure}
\begin{figure}[ht]
    \centering
    \includegraphics[width=\linewidth]{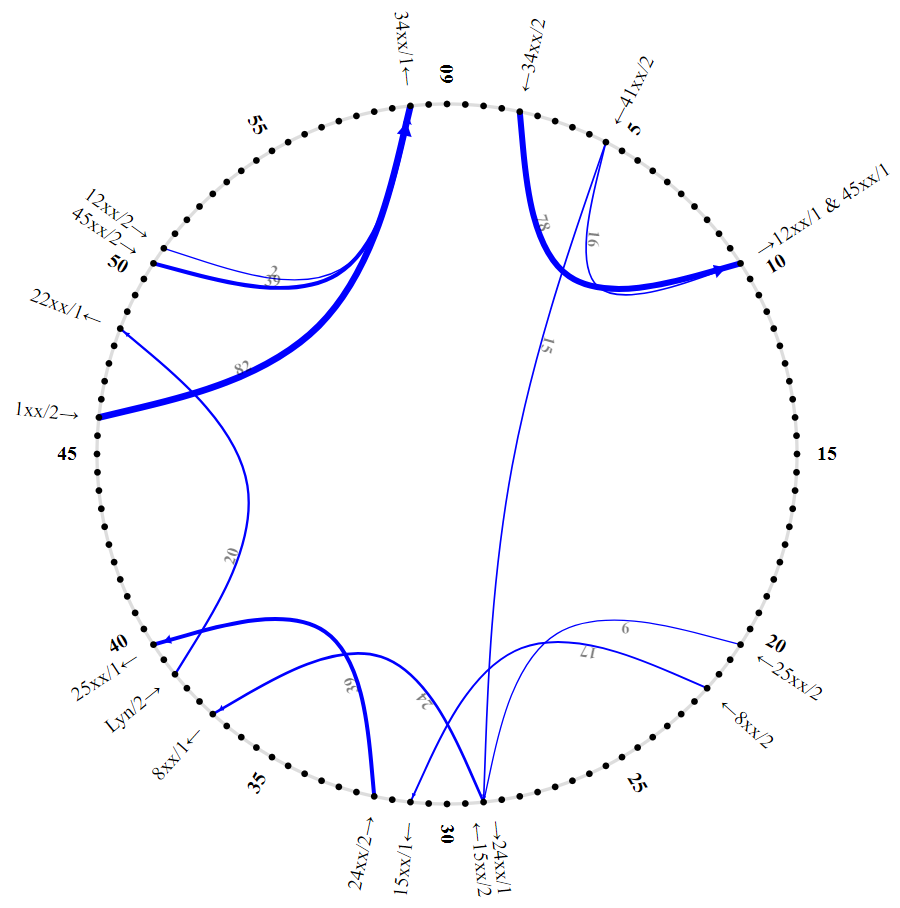}
    \caption{Example diagram of amount of passengers transferring between trains at K\o benhavn H during a rush hour}
    \label{fig:ChordKH}
\end{figure}





\end{document}